\documentclass[%
 aip,
 amsmath,amssymb,
 reprint,%
]{revtex4-1}

\usepackage[page, header]{appendix}
\usepackage{titletoc}
\usepackage{caption}

\usepackage{footnote}
\usepackage{amssymb}
\usepackage{amsmath}
\usepackage{graphicx}
\usepackage{subcaption}
\usepackage{mathtools}
\usepackage{bm}
\usepackage{color}
\usepackage{physics}
\usepackage{enumitem}
\usepackage[section]{placeins}
\usepackage{booktabs}
\usepackage{float}
\usepackage{caption}
\usepackage{siunitx} 
\usepackage{braket}

\usepackage{printlen}
\usepackage{pgf}

\usepackage[%
    colorlinks=true,
    pdfborder={0 0 0},
    linkcolor=blue,
    citecolor=blue,
    allcolors=blue
]{hyperref}

\begin{document}

\title{Interpreting Ultrafast Electron Transfer on Surfaces with a Converged First-Principles Newns-Anderson Chemisorption Function}

\author{Simiam Ghan}
\affiliation{Fritz Haber Institute of the Max Planck Society, Faradayweg 4-6, 14195 Berlin, Germany}
\affiliation{Chair for Theoretical Chemistry and Catalysis Research Center, Technical University of Munich, Lichtenbergstra\ss{}e 4, D-85747 Garching, Germany}
\author{Elias Diesen}
\affiliation{Fritz Haber Institute of the Max Planck Society, Faradayweg 4-6, 14195 Berlin, Germany}
\author{Christian Kunkel}
\affiliation{Fritz Haber Institute of the Max Planck Society, Faradayweg 4-6, 14195 Berlin, Germany}
\author{Karsten Reuter}
\affiliation{Fritz Haber Institute of the Max Planck Society, Faradayweg 4-6, 14195 Berlin, Germany}
\author{Harald Oberhofer}
\email{harald.oberhofer@uni-bayreuth.de}
\affiliation{Department of Physics and Bavarian Center for Battery Technologies, University of Bayreuth, Bayreuth, Germany.}

\begin{abstract}
We study the electronic coupling between an adsorbate and a metal surface by calculating tunneling matrix elements H$_{\text{ad}}$ directly from first principles. For this we employ a projection of the Kohn-Sham Hamiltonian upon a diabatic basis using a version of the popular  Projection-Operator Diabatization approach. An appropriate integration of couplings over the Brillouin zone allows the first calculation of a size-convergent Newns-Anderson chemisorption function, a coupling-weighted density of states measuring the line broadening of an adsorbate frontier state upon adsorption. This broadening corresponds to the experimentally-observed lifetime of an electron in the state, which we confirm for core-excited $\mathrm{Ar}^{*}(2{p}_{3/2}^{\ensuremath{-}1}4s)$ atoms on a number of transition metal (TM) surfaces. Yet, beyond just lifetimes, the chemisorption function is highly interpretable and encodes rich information on orbital phase interactions on the surface. The model thus captures and elucidates key aspects of the electron transfer process. Finally, a decomposition into angular momentum components reveals the hitherto unresolved role of the hybridized $d$-character of the TM surface in the resonant electron transfer, and elucidates the coupling of the adsorbate to the surface bands over the entire energy scale.
\end{abstract}

\maketitle

\section{Introduction}

The electronic coupling $\text{H}_{\text{ad}}$ between an adsorbate and a surface is a central quantity in many technologically important physical processes, including photochemistry, photovoltaics~\cite{Newton1991,Newton1996,Newton1997}, and heterogeneous
catalysis~\cite{Norskov1989,Hammer1995, Norskov1995,Schmickler2012}, not to mention its place in the very theory of the chemical bond~\cite{Hoffmann1988}. The coupling of an adsorbate to the $d$-band of transition metal surfaces has, for example, famously served as a descriptor to understand catalytic reactivity trends for a wide array of important chemical reactions~\cite{Pedersen2007,Pedersen2022,Norskov1989,Hammer1995,Norskov1995,Bligaard2008,Nilsson2008,Gross2009,Norskov2023}. Electronic coupling likewise plays an important role in the interpretation of experimental studies of charge transfer (CT) between adsorbates and surfaces, where it is invoked to understand e.g.~non-adiabatic excited-state dynamics~\cite{Persson1980, Saalfrank2006, Menzel2012},
desorption induced by (multiple) electronic transitions (DIET/DIMET)~\cite{Menzel1964, Misewich1992}, scanning-tunneling microscopy (STM)~\cite{Bardeen1961,Tersoff_Hamann_1984, TersoffHamann1983, Chen1990,Chen1992,Chen2007}, or pump-probe and core-hole clock spectroscopies. 

Given the prevalence of electronic couplings in theoretical descriptions, there has been a high interest in determining them with computational methods~\cite{Domcke1981_original_BD_paper,Newton1991,Norskov1989}, including, but not limited to, \textit{ab initio} methods~\cite{Norskov1989,Oberhofer2017}.
Coupling matrix elements $\text{H}_{\text{ad}}$ themselves are not physical observables though, and they thus evade direct comparison to experiment~\cite{Moon2008}. Furthermore, their definition in the theory is in general non-unique, complicating efforts to calculate them and compare alternative approaches~\cite{Newton1996,Voorhis2010,Ghan2020}. A careful interplay between theory and experiment is therefore necessary for a meaningful quantitative determination of these coupling parameters, which are most often approximated~\cite{Norskov1989,Maurer2021,Pedersen2014} and used only qualitatively.

\begin{figure}[ht!]
    \centering
\includegraphics[width=0.5\textwidth,scale=1, angle=0]{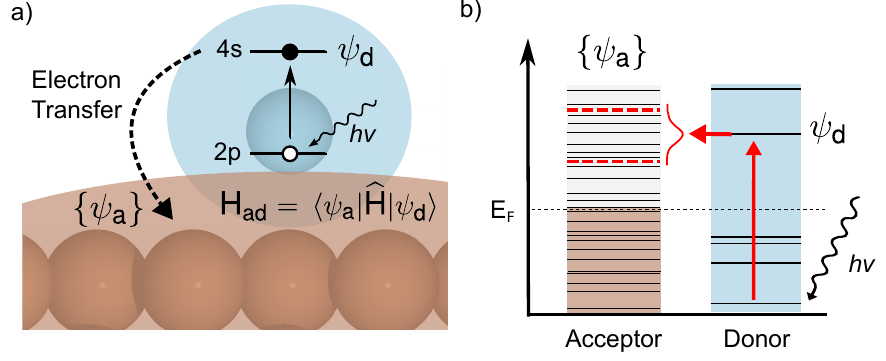}
\caption{a) Diagram of the core-excited electron transfer system of an Ar atom adsorbed on a transition metal surface. Ar is thereby optically core-excited into a 4s donor state $\psi_\text{d}$ which can then decay into the spectrum of surface acceptor states $\lbrace\psi_\text{a}\rbrace$ b) Schematic energy diagram of the process.}
    \label{fig:diagram}
\end{figure}

Of the many experimental setups where electronic couplings play a pivotal role, CT experiments provide a most natural context in which to test and validate a quantitative theory of electronic coupling. For example, experiments based on core-hole clock spectroscopy have measured the ultrafast electron transfer (ET) dynamics of simple core-excited adsorbates, such as  $\mathrm{Ar}^{*}(2{p}_{3/2}^{\ensuremath{-}1}4s)$, on transition metal surfaces to high precision~\cite{Menzel_1998,Menzel_2008,Menzel_2009,Menzel1988,Wurth2002,Feulner_2005,Feulner_2007,Feulner_2010,Feulner_2011,Feulner_2011,Feulner_2014,Feulner_2015,Feulner2016}. As illustrated in Figure \ref{fig:diagram}a, the probe atoms are thereby physisorbed to ultra-cold surfaces and optically excited. The thus resulting excited donor state -- labelled by the index d -- then decays on femtosecond time scales into the unoccupied acceptor states -- labelled with an a -- of the surface. A schematic of the energetics of such a process exemplified here by adsorbed Argon is given in Figure \ref{fig:diagram}b. Interpretation of such experiments usually proceeds through theoretical models of various levels of complexity \cite{Raseev2001,Echenique2007}. From a theoretical perspective, systems like this provide an ideal test-bed for electronic coupling models. First, with their ET timescales in the regime of a few femtoseconds, nuclear dynamics plays little to no role in the mechanism. Together with the fact that Ar only physisorbs at such surfaces, this greatly reduces the number of possible influences an ET theory needs to consider \cite{Oberhofer2017}. Yet, even such simple systems can show quite a complex ET behavior, strongly influenced by the type and density of the surface acceptor states and their coupling to the transferred electron~\cite{Menzel_2008,Wurth2002}. With the nowadays achievable experimental resolutions at the attosecond scale~\cite{Feulner_2005}, they thus offer fundamental insights into the ET process and provide a valuable test case for theories of electron transfer. 

Early experiments~\cite{Menzel2000,Wurth2002} and subsequent theoretical work~\cite{Raseev2001,Gauyacq2006,Borisov2004,Echenique2007,Muller2020} have explained measured (and simulated) ET lifetimes primarily in terms of the surface electronic band structure, invoking the projected band gap which is present in the dispersive $sp$-bands on many transition metal surfaces. These models thus centered solely on a coupling to the $sp$-type states, which is easily approximated using expressions common in STM simulations and which references their free-electron-like character~\cite{Bardeen1961,Tersoff_Hamann_1984,TersoffHamann1983,Chen2007}. Even though neglecting the possible role of the $d$-band, these models have succeeded in explaining aspects of the ET process, including its spin-dependence on magnetic surfaces~\cite{Muller2020} and the observed energy dependence of the ET lifetimes~\cite{Echenique2007}. The latter dependence results simply from the relative position of the resonance within the projected $sp$ gap, which reduces the density of available acceptor states and hinders (resonant) ET with high $\textbf{k}_{\perp}$ momentum (perpendicular to the surface).
This simple \textit{sp}-band model of ET is generally well motivated for Ar adsorbates on transition metal surfaces. Argon atoms physisorb more than 3\,\AA{} above the surface, and their states should thus overlap more strongly with diffuse and free-electron-like $sp$-states than with more localized $d$-states. Furthermore, the resonance energy lies some 3 eV above the substrate's Fermi energy, and thus well above the typical metal $d$-band. Nevertheless, the limitations due to a complete neglect of possible $d$-band coupling were pointed out already by Menzel in 2000~\cite{Menzel2000,Wurth2002} and Gauyacq and Borisov in 2004 \cite{Borisov2004}. In these studies, a model based on $sp$-band arguments alone and neglecting a possible $d$-character of states, was thought to be inadequate for a comprehensive understanding of differences in ET lifetimes observed on different surfaces. First-principles simulations were therefore called for explicitly to resolve the coupling to the $d$-type states, which is more difficult to describe~\cite{Menzel2000, Wurth2002}.  While recent {\em ab initio} methods, based on density-functional theory (DFT) combined with surface Green's function techniques, have succeeded in accurate predictions of the ET lifetimes on a number of surfaces~\cite{Echenique2007,Feulner_2005,Muller2020}, the limited interpretability of these methods---which proceeds by the same $sp$-band model as above---has left the role of coupling to the $d$-band in ultrafast ET unresolved. 

In this work, our objective is twofold. First, we provide a first-principles model of ultrafast ET on metal surfaces which is based on the direct calculation of electronic couplings H$_{\text{ad}}$ between the adsorbate and all surface bands. By calculating electronic couplings explicitly, we aim for enhanced interpretability compared to other highly accurate first-principles models~\cite{Echenique2007,Feulner_2005,Muller2020}, which nevertheless rely on the $sp$-band approximation for their mechanistic interpretation. Our key tool for this is the chemisorption function, from the celebrated model of Newns and Anderson~\cite{Newns1969,Anderson1961,Grimley1967, Maurer2021}. The chemisorption function derives the energetic broadening of an adsorbate resonance explicitly in terms of the couplings to the surface. This broadening in turn corresponds to the lifetime of an electron in the state, and can be compared directly with experimental ET lifetimes, offering a crucial bridge between observable lifetimes and non-observable couplings. Second, by critically validating the scheme against resonant ET measurements of Argon on five transition metal surfaces, and by invoking rules for electronic couplings based on phase arguments from STM, we seek to provide a reliable and general protocol for calculating electronic couplings between adsorbate and substrate bands over the entire energy scale, noting again their potential for broader use.

In the following, we describe how we use the Projection-Operator Diabatization method (POD)~\cite{Kondov2007} to extract the Argon/surface electronic couplings from Kohn-Sham DFT. We build here on insights from our earlier work, which offers a variant, POD2GS~\cite{Ghan2020}, with improved basis set convergence properties.  We additionally build on earlier applications of POD which calculated electronic coupling for photoexcited systems on surfaces~\cite{Kondov2007,Jingrui2008,Jingrui2010,Jingrui2012,Jingrui2015,Futera2017,Prucker2013,Prucker2018,Pastore2015}. While these works limited themselves to nonperiodic cluster models, or periodic models within a $\Gamma$-point approximation, we demonstrate within our scheme that it is possible and indeed necessary to consider couplings throughout the full Brillouin zone to achieve a convergent, and thus physically interpretable, chemisorption function of the surface.

\section{Theory}

\subsection{Newns-Anderson Chemisorption Function}

The chemisorption function~\cite{Newns1969} describes the energetic broadening or linewidth of an adsorbate frontier state $\ket{\psi_{\text{d}}}$ (donor) as it interacts with a continuum of states $\{\ket{\psi_{\text{a}}}\}$ (acceptors) 
on a surface, cf. Figure \ref{fig:diagram}: 
\begin{eqnarray}
\nonumber
\Delta (\epsilon) &=& \pi \sum_{\text{a}}
|\bra{\psi_{\text{a}}} \widehat{\text{H}} \ket{\psi_{\text{d}}} |^2 \delta( \epsilon - \epsilon_{\text{a}})~ \\
&=& \pi \sum_{\text{a}}
|{\rm H}_{\rm ad}|^2 \delta( \epsilon - \epsilon_{\text{a}})~,
\label{FGR}
\end{eqnarray}
where $\widehat{\text{H}}$ is the Hamiltonian of the full system and $\epsilon_{a}$ the energy of the acceptor states. Its relation to CT processes is given, in a weak coupling regime, by the Golden Rule transition  for a donor state of energy $\epsilon_{\text{d}}$:
\begin{subequations}
\begin{align}
\Gamma_{\text{d} \to \text{a}}(\epsilon_\text{d}) = \frac{2}{\hbar}\Delta(\epsilon=\epsilon_{\text{d}})\; ,
\end{align}
with the lifetime then given by
\begin{align}
\tau_{\text d}(\epsilon_\text{d}) = \Gamma_{\text{d} \to \text{a}}(\epsilon_\text{d})^{-1} \; .  
\end{align}
\label{eq:rates}
\end{subequations}
The chemisorption function thus emerges as the half-width at half-maximum (HWHM) of a Lorentzian broadening of the energy level $\epsilon_{\text{d}}$. It is also referred to as the weighted density of states (WDOS)~\cite{Schiffer2022}, as well as a hybridization function~\cite{Bahlke2021}, which emphasizes that it can in principle be evaluated at any energy $\epsilon$. 
Of course, the relation to the lifetime of the donor state strictly holds only for $\Delta (\epsilon_{\text{d}})$, but approximately one can also evaluate the chemisorption function for other energies, for example at the experimental resonance energy rather at a computed donor state energy $\epsilon_{\text{d}}$.

Importantly, the chemisorption function thus allows to approximately
explore and predict the lifetime as a function of the energy of the donor state. To adapt its definition for practical use in a periodic DFT supercell model for the extended metal surface, two modifications are needed. First, due to the finite nature of periodic slab models and the employed basis set, only a discrete quasicontinuum of $N$ levels is present. We therefore replace the Dirac $\delta$-functions in eq.~(\ref{FGR}) by artificial broadening functions, $g_{\sigma}(\epsilon - \epsilon_{\text{a}})$, to give $\Delta(\epsilon)$ a smooth and continuous behavior. Specifically, we choose a Lorentzian distribution with HWHM $\sigma$=0.2 eV, and demonstrate in the Supporting Information (SI) that the determined lifetimes do not depend on the choice of the latter parameter. Second, addressing the Bloch description of the electronic states in supercell models, the explicit state summation in eq.~(\ref{FGR}) needs to be replaced by an appropriate integration over the system's Brillouin zone (BZ) \cite{Kittel1987quantum}. Already including the Lorentzian smoothing, we correspondingly arrive at {\bf k}-resolved chemisorption functions
\begin{equation}
\Delta_{\textbf{k}}(\epsilon) = \pi\sum_{\text{a}}^{N} |\text{H}_{\text{ad},\textbf{k}} |^2 g_{\sigma}(\epsilon_{\textbf{k}} - \epsilon_{\text{a},\textbf{k}} )~,
   \label{gamma_k}
\end{equation}
with the full chemisorption then resulting from BZ integration~\cite{Marcus2000}
\begin{equation}
\Delta(\epsilon) = \frac{1}{\Omega_{\text{BZ}}}     \int_{\Omega_{\textbf{BZ}}} 
\Delta_{\textbf{k}}(\epsilon)
d\mathbf{k}~.
\label{gamma_k_BZ}
\end{equation}
In practice, this integral is performed by summing over the {\bf k}-points with weights $w_\textbf{k}$ of a regular Monkhorst-Pack grid in the irreducible BZ~\cite{Monkhorst1976}.

\subsection{Diabatization}

The donor $\ket{\psi_{\text{d}}}$ and acceptor $\{ \ket{\psi_{\text{a}}}\}$ states are the key inputs to compute the three quantities determining the chemisorption function:
\begin{subequations}
\begin{eqnarray}
\epsilon_{\text{a},\mathbf{k}} &=& \mel{\psi_{\text{a},\mathbf{k}}}{\widehat{\text{H}}_{\mathbf{k}}}{\psi_{\text{a},\mathbf{k}}}
\label{equ:acceptor_energy} \\
\epsilon_{\text{d},\mathbf{k}} &=& \mel{\psi_{\text{d},\mathbf{k}}}{\widehat{\text{H}}_{\mathbf{k}}}{\psi_{\text{d},\mathbf{k}}}
\label{equ:donor_energy} \\
\text{H}_{\text{ad},\mathbf{k}} &=& \mel{\psi_{\text{a},\mathbf{k}}}{\widehat{\text{H}}_{\mathbf{k}}}{\psi_{\text{d},\mathbf{k}}}.
\label{equ:k_coupling1}
\end{eqnarray}
\label{eq:matelements}
\end{subequations}
Newns~\cite{Newns1969} originally defined $\ket{\psi_{\text{d}}}$ and $\{ \ket{\psi_{\text{a}}}\}$ simply as the eigenfunctions of non-interacting (isolated) donor and acceptor fragments. We will term this the (frag$_d$,frag$_a$) diabatic basis, consisting in the present application case of the eigenstates of an infinitely periodic isolated Argon layer and an isolated metal surface, both described in a periodic-boundary condition supercell. 

For a more quantitative description of the ET process, we instead here rely on a diabatization procedure, which accounts for some weak hybridization between the donor and acceptor system. Specifically, we will draw on our previously described Projection-Operator Diabatization approach, which is based on the partitioning and block-diagonalization of the Kohn-Sham DFT Hamiltonian of the full interacting system (POD2), and the subsequent orthogonalization of the diabats via a Gram-Schmidt (GS) procedure (POD2GS)~\cite{Ghan2020}.

For ET from weakly physisorbed Argon to a metal surface, we thus generate a POD2 diabatic spectrum $\{\ket{\psi_{\text{d}}}\}$ for the Argon donor by extracting and diagonalizing its block in the Hamiltonian of the combined system (Figure~\ref{fig:gammapoint}b). This POD2 spectrum is found to contain a suitable Argon 4$s$-like lowest unoccupied molecular orbital (LUMO) state, and we refer to this donor state as POD2$_d$. Keeping to the original spirit of Newns, this is combined with the eigenstates of the isolated surface to yield the (POD2$_d$,frag$_a$) diabatic basis. Note that a full POD2-based approach where also the acceptor is described in the interacting picture (POD2$_d$,POD2$_a$) would not show appreciable differences to the (POD2$_d$,frag$_a$) representation, due to the acceptor being only very weakly influenced by the donor. 

In neither the (frag$_d$,frag$_a$) nor the (POD2$_d$,frag$_a$) basis, the donor and acceptor states are necessarily orthogonal to each other, as they should be for
a numerically stable and formally correct
value of  $\text{H}_{\text{ad}}$ according to tight binding theory~\cite{Ghan2020,Valeev2006,Baumeier2010}. To remove any finite overlap
\begin{equation}
\text{S}_{\text{ad},\mathbf{k}} = \braket{   \psi_{\text{a},\mathbf{k}}  |   \psi_{\text{d},\mathbf{k}}   } \quad ,
\label{k_overlap}
\end{equation}
we therefore GS orthogonalize each donor-acceptor state pair \cite{Ghan2020}. As a consequence,
the acceptor acquires a small orthogonalization tail, and the acceptor states are no longer orthogonal with respect to each other. For the present application, the effect of this orthogonalization is relatively small though. 
Consistent with earlier work \cite{Gauyacq2006}, the POD2$_d$ donor state has a slightly hybridized 4$s$4$p_z$ nodal structure (cf.~Figure \ref{fig:gammapoint})
and is already nearly orthogonal to the surface as a response to Pauli repulsion interactions. The orthogonalization leads thus only to a slight reduction of the electronic couplings, similar to the findings in earlier studies of molecular systems  \cite{Ghan2020,Valeev2006}. In the following we shall refer to the two orthogonalized diabatic basis sets as (frag$_d$,frag$_a$)GS and (POD2$_d$,frag$_a$)GS and compare the effect of the diabatization on the electronic couplings and ET lifetimes below. 

\subsection{Connection to Alternative Approaches}
\label{ssec:alternatives}

A key feature of the POD2 diabatization approach is that it can be performed at arbitrary points in {\bf k}-space, thus giving full access to the {\bf k}-resolved chemisorption functions $\Delta_{\bf k}(\epsilon)$ and allowing to properly perform the BZ integration in eq.~(\ref{gamma_k_BZ}). Note, though, that this feature is not unique to the POD2 method, but could quite simply be included in other schemes as well. For example, other Hamiltonian fragmentation methods such as fragment orbital DFT (FO-DFT) \cite{Siebbeles2003,Li2007JCP} or frozen-density embedding \cite{Wesolowski1993JPC,Pavanello2011JCP}
could be modified exactly analogously to the POD2 variants. Indeed, the (frag$_d$,frag$_a$)GS basis is here computed with such a modified FO-DFT~\cite{Schober2016}.
Similarly, though possibly with mildly more effort, other diabatization approaches such as constrained DFT \cite{Voorhis2010}
or the analytic overlap method \cite{Gajdos2014JCTC}
can be adapted to allow a proper integration over reciprocal space. Yet, in practice {\bf k}-point sampling has rarely been used in the context of adsorbate-substrate CT studies to date and, to the best of our knowledge, never been used to computing {\bf k}-dependent couplings from first principles  \cite{Tersoff_Hamann_1984,Marcus2000}.
We believe the reasons for this to be two-fold. First, most of these methods were initially developed to study molecular systems~\cite{Oberhofer2017}, which are either non-periodic or which can be very well approximated through cluster models. The second reason is that even inherently periodic systems were in the past often either approximated through cluster models or simply treated at the $\Gamma$-point. In the present context and nomenclature, this neglect of appropriate BZ integration would correspond to approximating the chemisorption function as
\begin{equation}
\Delta(\epsilon) \approx \Delta_{\textbf{k}=\Gamma}(\epsilon) \quad .
\end{equation}
This can be further simplified by assuming a constant effective coupling between the donor and the acceptor states to ultimately reduce eq.~(\ref{gamma_k}) to 
\begin{equation}
\Delta_\text{FGR}=  \pi|\text{H}_\text{eff} |^2 \text{DOS}(\epsilon) \;,
\end{equation}
with $\text{DOS}(\epsilon)$ denoting the density of states of the acceptor. This is equivalent to the famous Golden Rule of Fermi \cite{Fermi1950nuclear,Kondov2007,Prucker2013} and allows the interpretation of excited state lifetimes in terms of the
DOS \cite{Menzel2000,Wurth2002}. Note, though, that this simple picture would not yield quantitative results for extended systems because the DOS does not converge with system size as it scales with the total number of electrons. Instead, one could replace the DOS with a localized version as in the Tersoff-Hamann approach~\cite{Tersoff_Hamann_1984,TersoffHamann1983}.

Similarly, the prevalent $sp$-band model of ET would correspond to simply setting to zero all couplings H$_{\text{ad},\mathbf{k}}$ to $d$-like states in the acceptor spectrum (e.g.~according to a Mulliken analysis). The resulting $sp$-WDOS differs only from the commonly applied $sp$-band models by computing the remaining couplings from first principles instead of approximating them, either trivially as in the popular Tersoff-Hamann scheme~\cite{Bardeen1961,Tersoff_Hamann_1984,TersoffHamann1983,Chen2007}, or as the overlap of specifically constructed donor and acceptor states~\cite{Echenique2007,Muller2020}.
Note that such an approach allows the interpretation of ET lifetimes in terms of the relative position of the resonance energy in the surface-projected band gap in the $sp$-states, yielding a simple descriptor for ET processes \cite{Muller2020,Echenique2007}. The rationale behind this descriptor is that the $sp$-band gap may influence lifetimes both by reducing the total DOS of acceptor states at $\textbf{k}=\Gamma$, and by reducing the overall coupling by forcing transfer to states with increasingly large $\textbf{k}_{\parallel}$ values \cite{Muller2020,Echenique2007}. These, for a fixed energy, have less momenta in the $\textbf{k}_{\perp}$ direction and hence less overlap  and coupling with the adsorbate.  

Focusing solely on CT, all of the above methods, including our WDOS approach, give direct access to excited state lifetimes. Similar results can, of course, be achieved by direct time propagation of the initial state \cite{Raseev2001,Borisov2004,Gauyacq2006}. Most such prior approaches employed model Hamiltonians in the construction of the time propagator and indeed our diabatic Hamiltonian could be similarly employed. 

Finally, we here compute the terms in eq.~(\ref{gamma_k_BZ}) from a slab model of the acceptor system, cf.~section \ref{sec:compDetails} below. While slab models have found numerous applications in surface science their finite representation of, in principle, semi-infinite metals might lead to artifical gaps in the acceptor spectrum~\cite{Muller2020}. In our approach this is addressed by converging the results with respect to slab depth and by introducing a small broadening in eq.~(\ref{gamma_k}). Given that only surface states are expected to show non-zero coupling to the donor, this approach seems justified. Indeed our results 
(cf.~Fig.~S3 in the SI) 
show the WDOS to be well converged already at 4-layered slabs.
There is, however, another way to include the effects of the continuum on a material's surface states. Using surface Green's functions constructed from a Hamiltonian combining bulk and slab DFT calculations, Echenique and co-workers \cite{Sanchez2007,Echenique2007} directly computed resonance lifetimes \cite{Muller2020}, with only the smallest of artificial broadenings for numerical reasons. A similar semi-infinite Hamiltonian could potentially be used to compute the couplings H$_{\text{ad},\mathbf{k}}$.

\section{Computational Details} 
\label{sec:compDetails}
We apply our chemisorption-function based scheme to low-coverage models of Argon monolayers on five low-index transition metal surfaces. Namely, the magnetic Fe(110), Co(0001) and Ni(111), as well as the non-magnetic Ru(0001) and Pt(111) surfaces. The extended surfaces are described in periodic boundary supercell models, comprising slabs consisting of eight layers and a vacuum region in excess of 40\,{\AA}. 
Ar is adsorbed at the metal surfaces' top position on one side of the slab such that there is one adsorbate per surface unit-cell. At the employed large surface unit-cells detailed in 
Fig.~\ref{fig:gammapoint}, 
this then describes a dilute overlayer with lateral Ar-Ar distances exceeding 8\,{\AA}.

The electronic structure of all systems is described at the DFT level, using the Perdew-Burke-Ernzerhof (PBE) functional~\cite{PBE} with Tkatchenko-Scheffler~\cite{vdW} dispersion correction. All calculations are performed with the all-electron DFT package FHI-aims~\cite{Aims2009,Aims2013}. Tier 1 numeric atomic orbital basis sets are employed for all metals, a Pople 6-311+G** valence triple-$\zeta$ basis set from Basis Set Exchange \cite{BSE2019} is used for Argon. Note that this split basis approach is necessary because while standard FHI-aims basis sets are able to very accurately represent the electronic structure of the metallic surfaces, they would yield erroneous unoccupied states of the Ar adsorbate. Especially, the crucial LUMO state would have shown a wrong level alignment and orbital geometry. The chosen triple-$\zeta$ Ar basis fully remedies this as depicted in Fig.~\ref{fig:gammapoint}.
Real-space quantities are represented on FHI-Aims' tight integration grids while BZ integration is performed on a (4$\times$4$\times$1) regular $\textbf{k}$-point grid \cite{Monkhorst1976}. The Ar adatom and the topmost four metal layers are fully relaxed until residual forces fall below 5\,meV/{\AA}.
A half core-hole is then included on Ar to simulate the effect of the core-excited state~\cite{Mitch_2019,Leetmaa_2010}. 

For the calculation of the more delicate electronic couplings, a single-point calculation is finally performed.
The basis set setting of the topmost metal layer is increased to tier 4. Note that all other metal layers are left at tier 1 and the less crucial integration grids were used at light settings for reasons of computational efficiency. 
The SCF-converged Hamiltonian and overlap matrices are printed at each point on a $(12 \times 12 \times 1)$ k-grid, and diabatization routines are performed in an external progam 
based on SciPy \cite{SciPy}. Consistent with the employed low-coverage model and the corresponding excitation of the quasi-isolated initial core electron to arbitrary momenta, the wavepacket of the excited Ar$4s4p_z$ state is assumed to be uniformly distributed in k-space.

As demonstrated in 
Figs. S1, S9 and S10
of the SI, the resulting chemisorption function and the lifetimes derived from it are fully converged with respect to the employed finite k-grid, the Lorentzian smearing $\sigma$ and the metal basis set.
Unfortunately, convergence with respect to the Ar basis set is not that straightforward. Addition of further diffuse functions will increase the basis set superposition. Any such participation of Argon basis functions in describing the slab density will lead to the appearance of a ghost state upon partitioning and diagonalization of the Argon block in the POD2 scheme. This state at 1.6\,eV below the Fermi level is already seen in the $\Gamma$-point in Fig.~(\ref{fig:gammapoint}c) at the triple-$\zeta$ basis set and fortifies for larger Ar basis sets. We find such larger basis sets to also decrease the $4s4p_z$ hybridization of the true donor state. To avoid these problems, we therefore stick to the triple-$\zeta$ Ar basis set, which yields a comparable hybridization for the Ar@Fe(110) system as found in earlier work \cite{Gauyacq2006,Borisov2004,Echenique2007, Muller2020}. An in-depth analysis of the influence of the Ar-basis can be found in the SI.

We also note that, in direct opposition to earlier results \cite{Echenique2007}, the k-grid integrated chemisorption function converges well with slab depth. As depicted in 
Figs. S3 and S4
of the SI, already 4 layers of metal atoms show a mostly converged WDOS with no noticeable improvement beyond 8 layers.

Finally, we also examined the convergence of the WDOS with donor coverage. As mentioned above, we target a dilute limit in our work, while in the experiment Ar actually forms a dense monolayer. Yet, at the small fluences of incidence photons only a small fraction of Ar atoms will be excited at any one time. Thus, the excited donor states will essentially be localized and dilute in real space, rather than forming an excited band in the Ar overlayer. 
As depicted in 
Fig. S6
of the SI, we find a small variation of computed lifetimes going from the c(2x6) coverage used here to the dilute p(5x5) overlayer, similar to earlier studies~\cite{Echenique2007}, which however does not significantly influence resulting trends.

\section{Results}

\subsection{Chemisorption function for Ar at five transition metal surfaces}

\subsubsection{$\Gamma$-point results}

\begin{figure*}[ht!]
    \centering 
    \includegraphics[width=\textwidth,scale=1, angle=0]{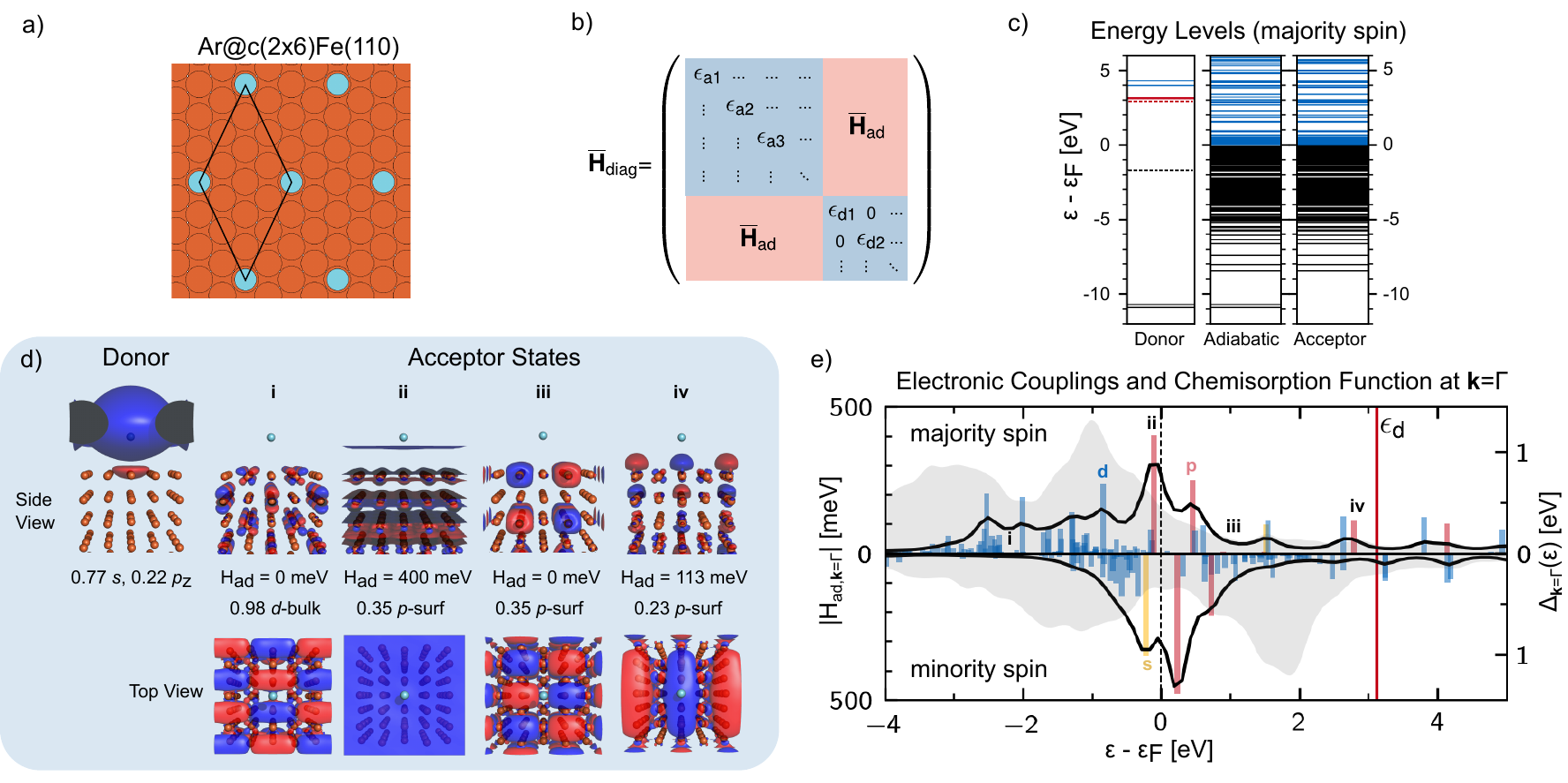}
\caption{Illustration of the diabatization procedure to obtain the chemisorption function $\Delta_{\textbf{k}=\Gamma}(\epsilon)$ at the $\Gamma$-point for the case of the low-coverage Argon model on ferromagnetic Fe(110) shown in a).
b) Schematic of the quasi block-diagonal representation of the Kohn-Sham Hamiltonian achieved upon projection onto the (POD2$_d$,frag$_a$)GS diabatic basis. 
c) Energy spectra of the majority-spin donor and acceptor diabatic blocks compared to the original adiabatic Kohn-Sham spectrum (occupied states in black, unoccupied states in blue). An Ar4$s$-like state (red solid line) emerges close to the experimental resonance (red dashed line) in the diabatic donor spectrum. Additionally, a ghost state resulting from basis set superposition appears at 1.6\,eV below the Fermi level (black dashed line).
d) Real-space representation of the POD2D Argon donor state, revealing its $4s4p_z$-type hybridization (isovalue 0.01 e$^{1/2}$\AA{}$^{-3/2}$), as well as of four select acceptor states indicated in panel e) (isovalues 0.05 (side views) and 0.02 (top view) e$^{1/2}$\AA{}$^{-3/2}$). In each case, the electronic coupling H$_{\rm ad}$ to the donor and the (Mulliken) character are indicated.
e) Electronic couplings H$_{\text{ad}}$ between the donor and all acceptor states are shown on the energy axis, colored according to the dominant character of the acceptor state in the metal slab
($s$ = yellow, $p$ = red, $d$ = blue). The resulting chemisorption function $\Delta_{\textbf{k}=\Gamma}(\epsilon)$, cf.~eq.~(\ref{gamma_k}), is shown as a black line. The predicted Ar donor state energy $\epsilon_\text{d}$ is marked with a red vertical line. For comparison, the $\textbf{k}=\Gamma$ DOS of the slab is shown light grey. 
}
 \label{fig:gammapoint}
\end{figure*}

To illustrate our approach, we first examine the properties of the chemisorption function at the $\Gamma$-point, 
$\Delta_{\textbf{k}=\Gamma}(\epsilon)$, 
for the low-coverage $c(2 \times 6)$ Ar/Fe(110) system. These results are summarized in Fig.~\ref{fig:gammapoint}. Upon projection onto the (POD2$_d$,frag$_a$)GS diabatic basis, the Kohn-Sham Hamiltonian nearly block-diagonalizes, with the acceptor block only approximately diagonal as a result of the GS orthogonalization procedure. The projection yields diabatic energy levels, which are comparable to the levels of the original adiabatic Kohn-Sham spectrum as shown in Fig.~\ref{fig:gammapoint}c.   
Most important for the ensuing discussion of adsorbate to surface ET, a donor state appears in the diabatic Argon spectrum at $\epsilon_\text{d} = 3.13$\,eV above the Fermi level. This is close to the experimental value of 2.97\,eV~\cite{Feulner_2014}
and we note that the half core-hole constraint is crucial to achieve this good alignment. As apparent from the real-space visualization in Fig.~\ref{fig:gammapoint}d 
this Ar donor state has a hybridized $4s4p_z$ character and thus contains a nodal plane parallel to the surface, consistent with earlier theoretical \cite{Gauyacq2006} and experimental \cite{Feulner2016} work.

Figure~\ref{fig:gammapoint}e shows the computed electronic couplings H$_{\text{ad},\textbf{k}=\Gamma}$ to all acceptor states in the diabatic basis (for comparison, the couplings at $\text{k}=\overline{\text{S}}$ are shown in 
Fig. S23 of the SI.
These couplings span a rather large range of strengths within each angular momentum channel. Most $d$-states show rather small couplings, seemingly supporting the prevalent $sp$-band model of ET. Yet, they are not zero, and in fact do show similar strengths as $s$- and $p$-like states in the energy region close to the donor level. In fact, the high symmetry of the Ar$4s4p_{z}$ donor state, as well as its position on the on-top site, allow to nicely understand these varying couplings in terms of textbook rules of orbital phase cancelation. Figure~\ref{fig:gammapoint}d illustrates this for four select acceptor states. Acceptor state i) is a bulk $d$-state with hardly any presence in the surface layer and a correspondingly vanishing coupling. State ii) is a $p$-like state which is in-phase with the donor orbital in the $\textbf{k}_{\parallel}$ direction, resulting in strong overlap and coupling. State iii) is the same state but out-of phase (antibonding) in the $\textbf{k}_{\parallel}$ direction, resulting in a cancellation of phase and a corresponding zero coupling. Finally, state iv) is an intermediate case with a corresponding intermediate coupling. 

In contrast to a mere (angular momentum projected) DOS the computed chemisorption function
$\Delta_{\textbf{k}=\Gamma}(\epsilon)$ shown in Fig.~\ref{fig:gammapoint}e appropriately accounts for these varying couplings, with its value 
$\Delta_{\textbf{k}=\Gamma}(\epsilon_{\text{d}})$ at the donor level energy then yielding the broadening for the Ar adsorbate and the concomitant ET lifetime. Coincidentally, the $\Gamma$-only WDOS displayed Fig.~\ref{fig:gammapoint}e already seems to provide, when evaluated at the donor resonance energy, lifetimes directly comparable to experimental results~\cite{Feulner_2014,Menzel2000,Nilsson_1996}.
Notwithstanding, we stress that the actual agreement here is in fact largely fortuituous. Primarily due to band folding, the WDOS merely evaluated in the $\Gamma$-point approximation is highly sensitive to the employed size of the periodic supercell. We illustrate this in 
Fig. S2
of the SI for a range of cell sizes. It is found that the chemisorption function within a $\Gamma$-point approximation only slowly converges with cell size, making it insufficient for the prediction of lifetimes at reasonable slab sizes. To really achieve a quantitatively converged chemisorption function, a proper integration over the BZ is indispensable.

\begin{figure}[ht!]
    \centering
    \begin{minipage}[b]{\linewidth}
\includegraphics[width=\textwidth,scale=1, angle=0]{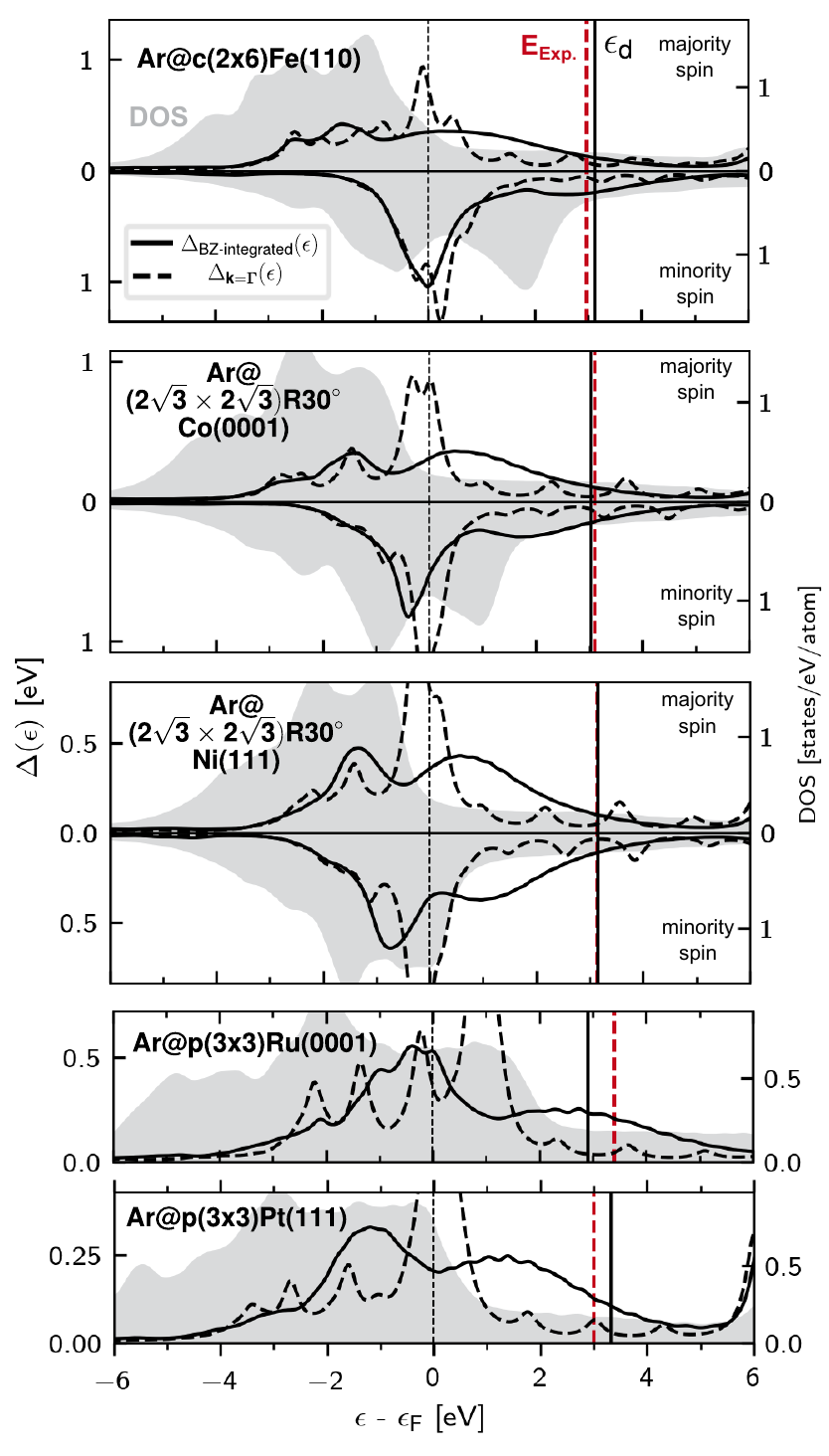}
\end{minipage}
\caption{
Brillouin-zone integrated chemisorption function (solid line) for the Ar$4s4p_{z}$ compared to the chemisorption function obtained at $\Gamma$-point (dashed line). The (POD2$_d$,frag$_a$)GS method is used.
The BZ-integration is performed on a (12$\times$12$\times$1) k-grid. Both functions contain a Lorentzian broadening of 0.2eV.  The experimental and calculated resonance energies are shown with vertical red and black lines, respectively. The DOS of the metal surface layer is shown for comparison.}
\label{fig:bz_vs_gammapoint}
\end{figure}

\subsubsection{Brillouin-zone integrated results}

Analogous to the $\Gamma$-point results in Fig. \ref{fig:gammapoint}e, we show in Fig. \ref{fig:bz_vs_gammapoint} now the Ar$4s4p_z$ chemisorption function on five transition metal surfaces after appropriate integration over the full Brillouin Zone (full lines), cf.~eq.  (\ref{gamma_k_BZ}).
To highlight the influence of BZ integration we depict the respective $\Gamma$-point results (dashed lines) alongside the $\mathbf{k}$-converged chemisorption functions.
Next to the noticeable differences between the two functions, we note that the BZ-integrated chemisorption function is actually largely independent of the slab size, unlike the $\Gamma$-point result. As depicted in 
Fig. S3
in the SI, we find the BZ-integrated WDOS to converge very well with slab depth, showing very little to no change beyond 4 layers. Furthermore, we find significant differences between the integrated and $\Gamma$-only WDOS' especially at the Argon donor energies. These, in turn, lead to noticeably different predictions for the lifetimes (cf.~eq. \ref{eq:rates}) of the core-excited Ar state.

\subsection{Lifetimes of core-excited Ar on surface systems}

\begin{figure}[ht!]
    \centering
  \includegraphics[width=0.4\textwidth,scale=1, angle=0]{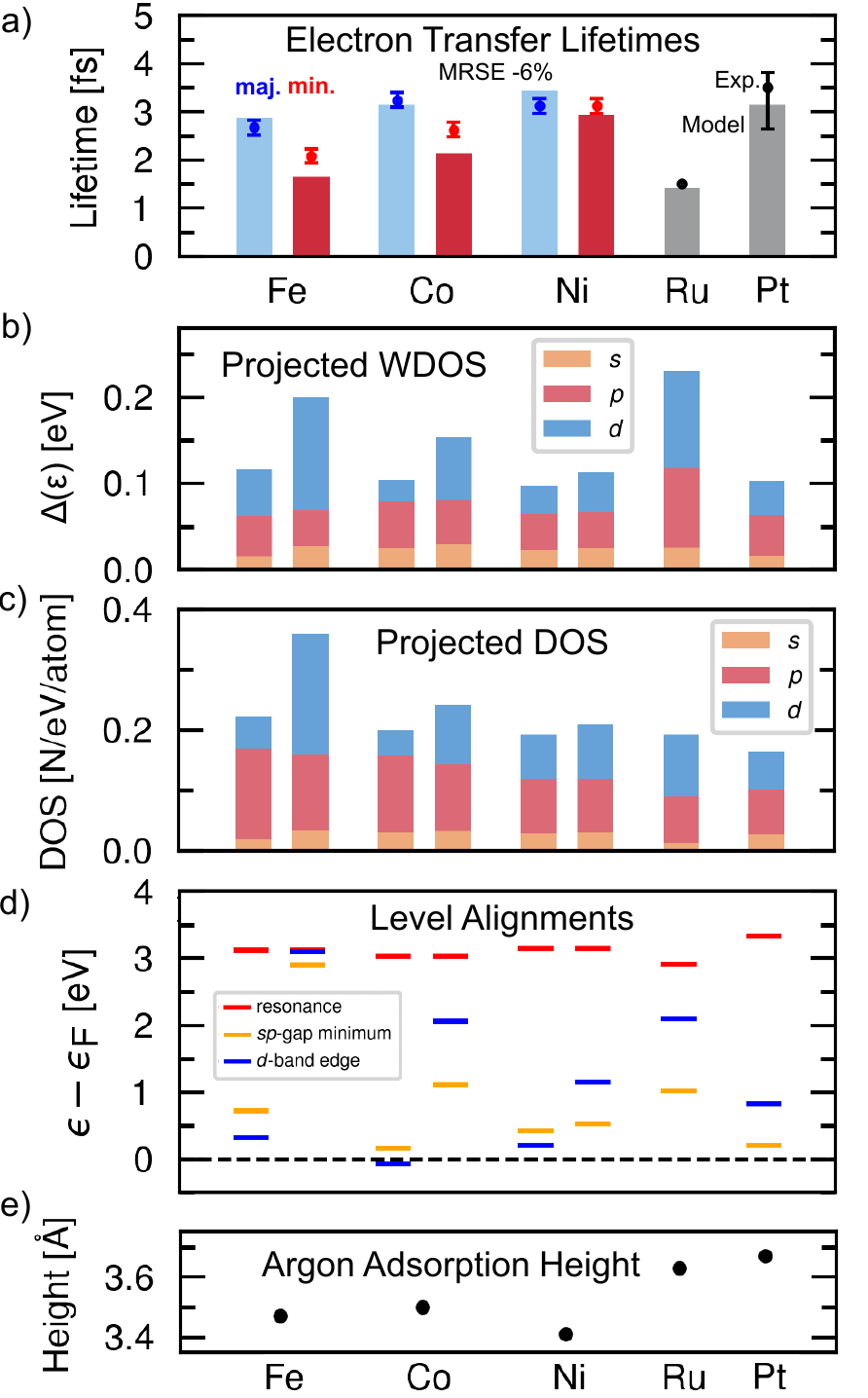}
\caption{ 
a) Calculated electron transfer lifetimes (vertical bars) compared to experimental values (dots, errorbars shown when reported). 
The lifetimes may be understood in terms of parameters describing the systems: b) the value of the chemisorption function (WDOS) and c) the DOS of the surface layer at the resonance energy, as well as their angular momentum decompositions, d) the alignment of the surface bands relative to the resonance, and e) the adsorption height of the Argon above the on-top site on each surface.  Correlations are analyzed in Fig.~\ref{correlations_mini}, and
Figs. S12 and S13 of the SI.
}
    \label{barplot_test3}
\end{figure}

The k-point integrated and converged chemisorption function now allows us to compute ET lifetimes (cf.~eq.~\ref{eq:rates}) comparable to experiment as illustrated in Fig. \ref{barplot_test3}a
for the transfer from $\mathrm{Ar}^{*}(2{p}_{3/2}^{\ensuremath{-}1}4s)$ to the five metal surfaces.
Additionally, the results are summarized in 
Table S5
of the SI.  Experimental lifetimes, lifetime errors (when reported), and resonance energies for ferromagnetic systems Fe(110), Co(0001) and Ni(111) were taken from the same publication \cite{Feulner_2014}, while those of Ru(0001) \cite{Menzel2000} and Pt(111) \cite{Nilsson_1996} are from separate, older works. For the core-hole measurement on Ar/Pt(111) \cite{Nilsson_1996}, no lifetime was reported. It can, however, be estimated with the same method as used in ref.~\cite{Menzel2000}. 
The calculated lifetimes have an overall mean relative signed error of \mbox{-6\%} and mean signed error of 
-0.14fs compared to the experimental values.
The qualitative and quantitative differences in lifetimes over the surfaces, and, for ferromagnetic systems, between spin channels, are captured excellently.

Having established the 
agreement of our approach with experiment, we
now examine the contributions of different acceptor states, specifically focusing on the angular momenta. For comparison, Figs. \ref{barplot_test3}b and c, respectively depict the WDOS and surface-DOS evaluated at the resonance energies and both resolved by angular momentum contributions. To this end, we employ a Mulliken analysis of both the DOS and WDOS 
(cf.~eq. S15 in the SI).
The angular-momentum components of the WDOS (which we shall call the projected-WDOS or pWDOS) thus contain the weights of each state \textit{by character}. The pWDOS can thus be seen as a projected DOS weighted by couplings, and thus shows the dominant character of the states which participate in the overall lifetime broadening
(see Fig. S14 of the SI). 
It does not necessarily show the character which is associated with strong coupling (indeed, coupling and character are nontrivially related, as was seen in Fig. \ref{fig:gammapoint}). 

Figure~\ref{barplot_test3}b shows that states which participate in the overall lifetime broadening have, on average, significant $d$ as well as $sp$-character.  This is of course consistent with the hybridized nature of states at the resonance energy.
However, we see upon closer examination that it is the changing $d$-character of the states which is found to decisively determine the differences in lifetimes over the surfaces and spin channels, while the $sp$-character is relatively unchanged.

Considering other popular modes of interpreting ET lifetimes, the DOS  depicted in Fig.~\ref{barplot_test3}c indeed bears some qualitative resemblances to the WDOS. Importantly, the WDOS, due to its inclusion of phase effects, captures more closely the variations among the surfaces, in closer agreement with the experimental lifetimes than the DOS. In particular, the DOS fails to capture the difference between the shortest (Ru) and longest (Pt) lifetimes. The WDOS, on the other hand, shows a much higher value on Ru than Pt, despite similar DOS values on the surfaces. As Argon adsorbs approximately at the same height on these two systems---cf.~Fig. \ref{barplot_test3}e, the large difference in lifetimes on Ru and Pt is strongly related to different electronic couplings, determined both by the spatial extent of the surface wavefunctions and specific phase effects. 

Finally, in Figure \ref{barplot_test3}d we depict the onsets of the $sp$-gap and the $d$-band together with the resonance energy. The $sp$-gap onset is thereby evaluated from the primitive band structures of all five surfaces 
(cf. ~Fig. S26 in the SI),
while the $d$-band edge is extracted from the materials' surface DOS at a cutoff of 0.2 [states/eV/atom]. Both, the proximity of the resonance to the $sp$-band gap onset and to the $d$-band appear to correlate well with the lifetimes. 
Yet, we note for example that for Fe, very different $sp$-bandgap onsets appear in the majority and minority spin channels, while the $sp$-DOS and $sp$-WDOS are nearly identical in both spin channels, indicating little influence of band gap alignment on the couplings or DOS of the $sp$-channel (indeed, the parabolic gap should have a nearly constant DOS similar to the 2D free electron gas). In contrast, the difference of the $d$-band edge between the spin channels corresponds to a large difference in both the $d$-band DOS and WDOS.

\begin{figure}[ht!]
    \centering
 \includegraphics[width=0.5\textwidth,scale=1, angle=0]{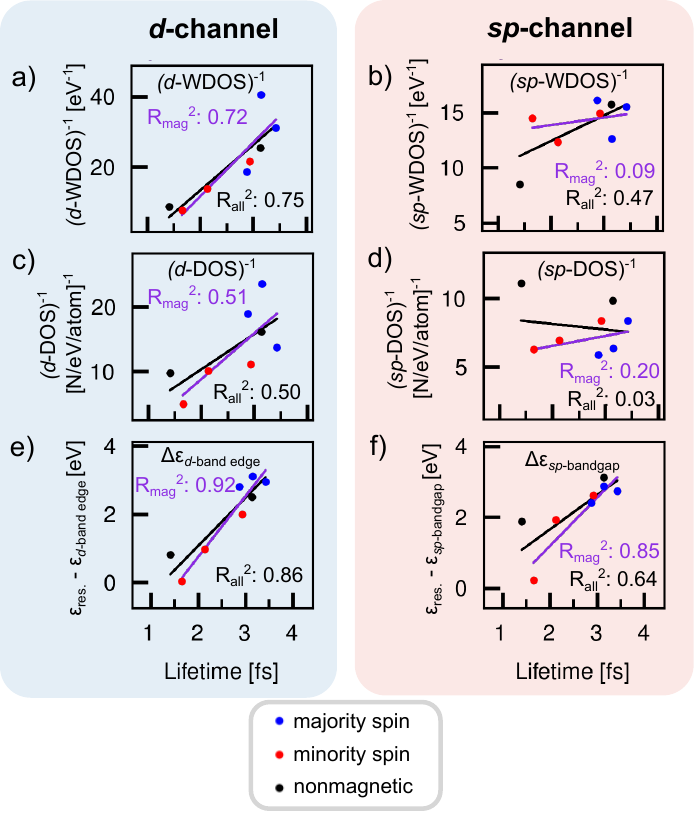}
\caption{ 
The predicted electron transfer lifetimes from Fig. \ref{barplot_test3} are tested for correlation with electronic structure properties of the surfaces. A linear regression is shown for all datapoints (black line) and seperately for magnetic systems (violet line).
The angular-momentum components of the WDOS are depicted: a) contains the WDOS weighted by the  $d$-character of states and  b) the WDOS weighted by the $sp$-character of states.
The projected DOS of the surface is shown: c) contains the DOS weighted by the $d$-character and d) the DOS weighted by the $sp$-character of states in the surface layer.
Finally, the proximity of the resonance energy to the edge of the $d$-band is depicted in e), with the onset energy of the $sp$-band gap depicted in d). 
}
    \label{correlations_mini}
\end{figure}

For a clearer analysis of such relationships, we correlate the ET lifetimes against features of the electronic structure in Figs. \ref{correlations_mini}. We find that the calculated lifetimes correlate strongly with (coupling-weighted) $d$-character of the acceptor states, and weakly with their $sp$-character. 
In Figs. \ref{correlations_mini}a,b, the (inverse) WDOS $d$-channel is seen to correlate strongly with the lifetimes, while the $sp$-channel of the WDOS correlates poorly. This correlation is even worse for the investigated magnetic systems. 
Furthermore, coming back to the simple Tersoff-Hamann picture, the ET rate should be proportional to the local DOS at the probe coordinate (approximated here as the DOS of the surface layer).
In Figs. \ref{correlations_mini}c,d, we correlate the lifetimes against the inverted $d$ and $sp$ components of the DOS, respectively. Thereby, we observe a very slight correlation between the lifetimes and inverse DOS of the $d$-channel, but none at all for the $sp$-DOS, consistent with the lack of phase information in this simple picture.
Additionally, while our predicted lifetimes correlate strongly with the proximity of the resonance to the $sp$-band gap onset as depicted in Fig. \ref{correlations_mini}f, we find them to correlate poorly with the $sp$-component of the DOS (Fig. \ref{correlations_mini}d and the $sp$-WDOS (Fig. \ref{correlations_mini}b). 
Instead, there is a strong correlation of the lifetimes with the $d$-band edge,
which may be explained by an increasing diffusivity of states with energy \cite{Echenique2007}, or simply by the increase in the $d$-band DOS magnitude at the $d$-band edge.
The energy of the $d$-states is captured in the $d$-band edge descriptor, and indeed strongly correlates with the $d$-band WDOS,
cf.~Fig.  S13c in the SI.

\begin{figure*}[ht!]
      \centering
\includegraphics[width=\textwidth,scale=1,angle=0]{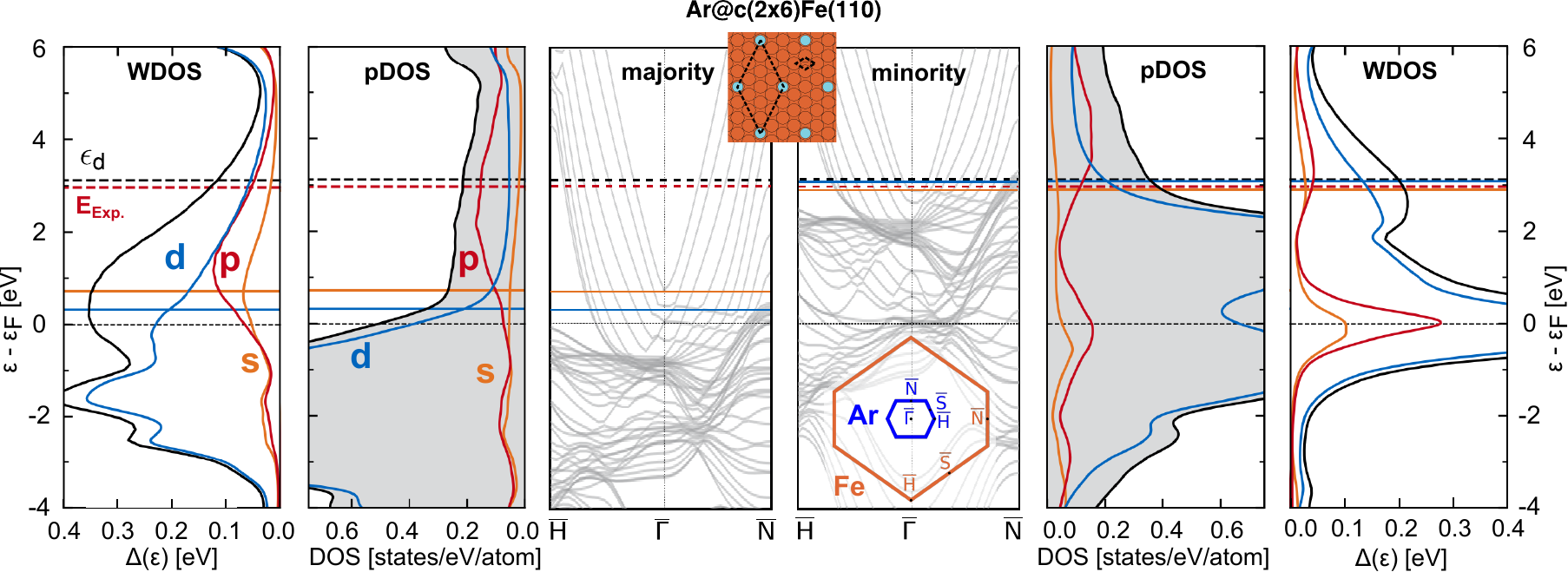}  
\caption{
The chemisorption function of the Ar/Fe(110) c(2x6) system is compared to the surface-projected DOS, and to the band structure of the surface primitive cell. The nontrivial relationship between the DOS and the WDOS arises due to the electronic couplings, which contain e.g. phase cancellation effects.
The $sp$-band gap onset, marked with an orange horizontal line, is associated with a local maximum in the chemisorption function followed by a monotonic decay, consistent with the role of the band gap identified in earlier work, and with the experimentally-measured energy dependence of the ET process. The $d$-band edge is marked with a blue line, and the experimental and predicted Ar$4s$ resonances are indicated. Other systems shown in 
Figs.~S15 and S26 
of the SI. 
}
    \label{band_plot_exploded}
\end{figure*}

We find that similar observations can be made when interpreting the energy dependence of the predicted electron transfer lifetimes. 
While the connection between the WDOS and excited-state lifetimes only strictly holds at the resonance energy cf.~Fig.~\ref{eq:rates}, assuming an unaffected nature of the donor state, one could use it to interpret experimental findings for other incidence energies. They show growing lifetimes for larger incidence energies beyond the resonance \cite{Keller1998,Menzel2000,Menzel_2008}. Furthermore, local maxima in the ET rates below the resonance have been observed experimentally \cite{Nilsson_1996} on Ar/Pt(111) and in accurate Green's Function-based simulations~\cite{Echenique2007} of Ar/Ru(0001). Both of these trends are  reproduced in the WDOS as depicted in Fig.~\ref{fig:bz_vs_gammapoint} and S18 of the SI. 
Theoretically, the energy dependence of lifetimes has alternatively been attributed to the declining DOS of the $d$-states (already in 1998 by Menzel and co-workers\cite{Keller1998,Menzel_1998}), or to the effect of the band gap in the $sp$-states originally proposed by Gauyacq and Borisov \cite{Borisov2004,Gauyacq2006,Echenique2007}.
Upon examination of the WDOS components over the energy scale, as shown in Fig. \ref{band_plot_exploded} for Ar/Fe, we find that the energy dependence at resonance arises from significant contributions from both $d$- and $sp$ components, combining the two explanations proposed earlier.  
The onset of the $sp$-band gap, which is marked with an orange line in Fig. (\ref{band_plot_exploded}), is associated with a local maximum in the WDOS, followed by a sharp monotonic decay which differs from the oscillatory behavior observed at lower energies.  The decay behavior is consistent with the decay of couplings to $sp$-type states with $\textbf{k}_{\parallel}$ in the gap region proposed earlier \cite{Echenique2007}. While we relegate a more detailed study of such dependencies to future work (the present model is based on the folded band structure, 
cf.~Fig.~S25 of the SI), 
we do observe this suppression of the $p$-channel WDOS in the band gap region of all systems 
(Fig.~S15 in the SI)
---against quite different behavior in the PDOS---indicating a systematic relationship between the WDOS and the band structure.

Finally, observing the nontrivial relationship of the DOS and WDOS over the energy scale, we can furthermore better understand the (lack of) correlation of the ET lifetimes with the $d$- and $sp$-channels established above.
For this, we may consider the nature of the couplings themselves.
States with strong $sp$ character, which have large spatial extent and presence on the surface, may nevertheless have vanishing couplings due to phase cancellation (cf.~Fig.~\ref{fig:gammapoint}). Their presence in the pWDOS, compared to the PDOS, is thus suppressed near the resonance in Fig. \ref{band_plot_exploded}.  The more localized $d$-bands, on the other hand, couple via their exponential tails and are thus less effected by phase cancellation.
Chen related \cite{Chen1990,Chen1992,Chen2007,Gross2011} the coupling matrix elements to the gradients, rather than the squared-modulus (density), of participating states. As such, we may anticipate qualitatively different coupling behaviors from the localized (high-gradient) $d$-band states compared the more diffuse (and uniform) free-electron-like $sp$-bands above the Fermi level.

These effects are meaningfully elucidated by constraining the donor to have purely 4s character, within the (frag$_d$,frag$_a$)GS method.  Upon removal of the 4$p_z$ hybridization, depicted in 
Figures S15 and S16
of the SI, we recover a much stronger relationship between the $d$-WDOS and $d$-DOS  over the full energy scale. Nonlinearity between the $sp$-DOS and $sp$-WDOS is meanwhile enhanced, confirming the increased vulnerability of these acceptor states to strong phase effects.

\section{Conclusions}
In this work, we examine the utility of the Newns-Anderson chemisorption function, also known as coupling-weighted DOS (WDOS), for the prediction and interpretation of ultrafast electron transfer lifetimes. Specifically, we revisit the case of core-excited $\mathrm{Ar}^{*}(2{p}_{3/2}^{\ensuremath{-}1}4s)$  atoms on five different transition metal surfaces, to find our WDOS results in very good agreement with experimental measurements.

We demonstrate that the WDOS converges well with system size, both lateral and with layer depth. A prerequisite for this convergence and indeed the agreement with experiment was found to be an appropriate sampling of each system's Brillouin zone. Using the tried and tested $k$-point scheme by Monkhorst and Pack to integrate the WDOS, we find electron transfer rates from the excited Argon to the metal surface to already converge for metal slabs of only 4 layers. 
This performance stands in contrast to the previously supposed inadequacy of finite slab models for the calculation of sensitive electron transfer properties, due to inherent finite size effects.   Previously, such effects were overcome by combining bulk and slab calculations and using surface Green's Functions techniques~\cite{Echenique2007,Sanchez2007,Kondov2007}.  Our model overcomes the finite size effects present in the $\Gamma$-point approximation simply with an integration over the Brillouin zone with a sufficiently dense k-grid and an energy broadening function, yielding a comparable accuracy to previous approaches for the present systems.

A critical component of the WDOS scheme is the calculation of electronic coupling matrix elements $\text{H}_{\text{ad}}$, for which a numerically stable diabatization procedure is necessary.   To this end, we apply the DFT-based Projection-Operator Diabatization scheme POD2~\cite{Ghan2020} to generate the diabatic basis of the Argon donor system, and use the eigenspectrum of the unperturbed surface as the diabatic basis of the acceptor (surface). 
The two systems are Gram-Schmidt orthogonalized, yielding the diabatic basis we term (frag$_{\rm a}$,POD2$_{\rm d}$)GS, and matrix elements are extracted by projecting the Kohn-Sham Hamiltonian onto this basis.

The resulting $\textit{ab initio}$ chemisorption function is found to be highly expressive, encoding rich information on orbital phase and symmetry, and yielding insight into the decay mechanism beyond a mere prediction of rates. A key advantage of this approach is its ability to explicitly capture complex phase cancellation effects over the full single-particle spectrum, hinting at an under-leveraged potential for diabatic representations of the Kohn-Sham Hamiltonian \cite{Voorhis2010}.
Across the studied systems, highly nontrivial and nonlinear phase cancellation effects are found to modulate the magnitude of the chemisorption function. This contrasts with the anticipated scaling between the electron transfer lifetimes and features of the $sp$-channel, which arises from a simplified free-electron-like view of couplings~\cite{Borisov2004,Gauyacq2006,Echenique2002,Echenique2007,Muller2020}, as openly recognized in earlier works.
Decomposing the WDOS by angular momentum contributions, we find phase cancellation to have less of an effect on the acceptor states with strong $d$-character, and it is the degree of $d$-hybridization of states which is then found to decisively modulate the electron transfer lifetimes, explaining their surface and spin dependence and contributing to their energy dependence.
By treating the DOS, the band structure, $\textit{and}$ the electronic couplings all in a consistent $\textit{ab initio}$ level of theory, the WDOS scheme thus resolves long-standing questions (formulated e.g.~by Menzel {\em et al.}~in 2000~\cite{Menzel2000,Wurth2002}, or later by Gauyacq and Borisov \cite{Borisov2004}) on the respective roles of the compact $d$-states versus the more spatially-extensive $sp$-states in the resonant ET process.

Note that the role of orbital phase cancellation in electron transfer has long been recognized in scanning tunneling microscopy~\cite{Chen1990,Chen1992,Gross2011,Repp2005}, where Chen's derivative rules were found essential to understanding the contrast mechanism beyond the (phase-less) Tersoff-Hamann model \cite{Tersoff_Hamann_1984,TersoffHamann1983}.
Our results for the ET lifetimes are thus perhaps unsurprising.  For magnetic systems, the $d$-band is, after all, that which is responsible for ferromagnetism and the asymmetry of the electronic structure.  
Furthermore, the observation of a relatively constant contribution of $sp$-type states to the lifetimes on different surfaces is also well in line with the prevailing view (e.g.~in heterogenous catalysis ~\cite{Pedersen2014,Bligaard2008,Pedersen2022}) that the coupling to the $sp$-band is a constant and largely independent of the substrate. 

The present model does, however, also show some departure from prevailing assumptions about the electronic coupling.
Noticeably, the phase modulation present in the $sp$-WDOS gives it a more rich and nontrivial dependence on the incidence energy, compared to the relatively structureless $sp$-projected DOS.
This is in contrast with the common wide-band approximation, which assumes a constant (energy-independent) coupling to the $sp$-bands~\cite{Subotnik2016, Subotnik2017, Subotnik2018,Anderson1961, Schmickler2012,Maurer2021,Pedersen2022}. Furthermore, the $d$-channel of the  chemisorption function, which shows a stronger resemblance to the $d$-band DOS (and indeed is often approximated as such~\cite{Pedersen2014}),
still shows nontrivial departures from the contour of the $d$-band DOS over the incidence energy scale.  

Taken together, all of the above observations suggest a much stronger role of phase effects in other approximative Newns-Anderson models,
such as those forming the staple description of e.g.~many chemical processes on surfaces~\cite{Norskov1989,Pedersen2014,Pedersen2022,Bukas2022,Kastlunger2022,Maurer2021,Bhattacharjee2016,Schmickler2012,Norskov2023}.

\begin{acknowledgments}
The authors thank Peter Feulner, Wolfgang Domcke, Alan Luntz and Niri Govind for useful and enlightening discussions. S.G.~was supported by the Deutsche Forschungsgemeinschaft (DFG, German Research Foundation) grant number RE1509/24-1. H.O.~recognizes support by the DFG under the Heisenberg program (grant no.~OB 425/9-1). This work was partially funded by the DFG under Germany’s Excellence Strategy (Grant No. EXC 852
2089/1-390776260).
\end{acknowledgments}

\section*{Supporting Information}
Equations S1-S6 clarify the Gram-Schmidt orthogonalization of the diabatic basis, S7-S10 the Mulliken analysis of states and S11-S16 the angular momentum decomposition of the WDOS.
Figures are provided to illustrate convergence of the chemisorption function with respect to: Brillouin zone sampling (S1,S2), slab depth (S3,S4),  and basis sets and monolayer coverage (S5-S10). Figures S11-S23 show $\textbf{k}=\Gamma$ and BZ-averaged diabatic properties comprising the calculation of ET lifetimes on the five slab systems.  Figures S24-S26 provide adiabatic DOS and band structures for comparison.


\bibliographystyle{apsrev4-2}
\bibliography{main}

\begin{thebibliography}{96}%
\makeatletter
\providecommand \@ifxundefined [1]{%
 \@ifx{#1\undefined}
}%
\providecommand \@ifnum [1]{%
 \ifnum #1\expandafter \@firstoftwo
 \else \expandafter \@secondoftwo
 \fi
}%
\providecommand \@ifx [1]{%
 \ifx #1\expandafter \@firstoftwo
 \else \expandafter \@secondoftwo
 \fi
}%
\providecommand \natexlab [1]{#1}%
\providecommand \enquote  [1]{``#1''}%
\providecommand \bibnamefont  [1]{#1}%
\providecommand \bibfnamefont [1]{#1}%
\providecommand \citenamefont [1]{#1}%
\providecommand \href@noop [0]{\@secondoftwo}%
\providecommand \href [0]{\begingroup \@sanitize@url \@href}%
\providecommand \@href[1]{\@@startlink{#1}\@@href}%
\providecommand \@@href[1]{\endgroup#1\@@endlink}%
\providecommand \@sanitize@url [0]{\catcode `\\12\catcode `\$12\catcode
  `\&12\catcode `\#12\catcode `\^12\catcode `\_12\catcode `\%12\relax}%
\providecommand \@@startlink[1]{}%
\providecommand \@@endlink[0]{}%
\providecommand \url  [0]{\begingroup\@sanitize@url \@url }%
\providecommand \@url [1]{\endgroup\@href {#1}{\urlprefix }}%
\providecommand \urlprefix  [0]{URL }%
\providecommand \Eprint [0]{\href }%
\providecommand \doibase [0]{https://doi.org/}%
\providecommand \selectlanguage [0]{\@gobble}%
\providecommand \bibinfo  [0]{\@secondoftwo}%
\providecommand \bibfield  [0]{\@secondoftwo}%
\providecommand \translation [1]{[#1]}%
\providecommand \BibitemOpen [0]{}%
\providecommand \bibitemStop [0]{}%
\providecommand \bibitemNoStop [0]{.\EOS\space}%
\providecommand \EOS [0]{\spacefactor3000\relax}%
\providecommand \BibitemShut  [1]{\csname bibitem#1\endcsname}%
\let\auto@bib@innerbib\@empty
\bibitem [{\citenamefont {Newton}(1991)}]{Newton1991}%
  \BibitemOpen
  \bibfield  {author} {\bibinfo {author} {\bibfnamefont {M.~D.}\ \bibnamefont
  {Newton}},\ }\href {https://doi.org/10.1021/cr00005a007} {\bibfield
  {journal} {\bibinfo  {journal} {Chem. Rev.}\ }\textbf {\bibinfo {volume}
  {91}},\ \bibinfo {pages} {767} (\bibinfo {year} {1991})}\BibitemShut
  {NoStop}%
\bibitem [{\citenamefont {Cave}\ and\ \citenamefont
  {Newton}(1996)}]{Newton1996}%
  \BibitemOpen
  \bibfield  {author} {\bibinfo {author} {\bibfnamefont {R.~J.}\ \bibnamefont
  {Cave}}\ and\ \bibinfo {author} {\bibfnamefont {M.~D.}\ \bibnamefont
  {Newton}},\ }\href {https://doi.org/10.1016/0009-2614(95)01310-5} {\bibfield
  {journal} {\bibinfo  {journal} {Chem. Phys. Lett.}\ }\textbf {\bibinfo
  {volume} {249}},\ \bibinfo {pages} {15 } (\bibinfo {year}
  {1996})}\BibitemShut {NoStop}%
\bibitem [{\citenamefont {Cave}\ and\ \citenamefont
  {Newton}(1997)}]{Newton1997}%
  \BibitemOpen
  \bibfield  {author} {\bibinfo {author} {\bibfnamefont {R.~J.}\ \bibnamefont
  {Cave}}\ and\ \bibinfo {author} {\bibfnamefont {M.~D.}\ \bibnamefont
  {Newton}},\ }\href {https://doi.org/10.1063/1.474023} {\bibfield  {journal}
  {\bibinfo  {journal} {J. Chem. Phys.}\ }\textbf {\bibinfo {volume} {106}},\
  \bibinfo {pages} {9213} (\bibinfo {year} {1997})}\BibitemShut {NoStop}%
\bibitem [{\citenamefont {N\o{}rskov}(1989)}]{Norskov1989}%
  \BibitemOpen
  \bibfield  {author} {\bibinfo {author} {\bibfnamefont {J.~K.}\ \bibnamefont
  {N\o{}rskov}},\ }\href {https://doi.org/10.1063/1.456679} {\bibfield
  {journal} {\bibinfo  {journal} {J. Chem. Phys.}\ }\textbf {\bibinfo {volume}
  {90}},\ \bibinfo {pages} {7461} (\bibinfo {year} {1989})}\BibitemShut
  {NoStop}%
\bibitem [{\citenamefont {Hammer}\ and\ \citenamefont
  {N\o{}rskov}(1995)}]{Hammer1995}%
  \BibitemOpen
  \bibfield  {author} {\bibinfo {author} {\bibfnamefont {B.}~\bibnamefont
  {Hammer}}\ and\ \bibinfo {author} {\bibfnamefont {J.}~\bibnamefont
  {N\o{}rskov}},\ }\href {https://doi.org/10.1038/376238a0} {\bibfield
  {journal} {\bibinfo  {journal} {Nature}\ }\textbf {\bibinfo {volume} {376}},\
  \bibinfo {pages} {238} (\bibinfo {year} {1995})}\BibitemShut {NoStop}%
\bibitem [{\citenamefont {Hammer}\ and\ \citenamefont
  {Nørskov}(1995)}]{Norskov1995}%
  \BibitemOpen
  \bibfield  {author} {\bibinfo {author} {\bibfnamefont {B.}~\bibnamefont
  {Hammer}}\ and\ \bibinfo {author} {\bibfnamefont {J.}~\bibnamefont
  {Nørskov}},\ }\href {https://doi.org/10.1016/0039-6028(96)80007-0}
  {\bibfield  {journal} {\bibinfo  {journal} {Surf. Sci.}\ }\textbf {\bibinfo
  {volume} {343}},\ \bibinfo {pages} {211} (\bibinfo {year}
  {1995})}\BibitemShut {NoStop}%
\bibitem [{\citenamefont {Santos}\ \emph {et~al.}(2012)\citenamefont {Santos},
  \citenamefont {Quaino},\ and\ \citenamefont {Schmickler}}]{Schmickler2012}%
  \BibitemOpen
  \bibfield  {author} {\bibinfo {author} {\bibfnamefont {E.}~\bibnamefont
  {Santos}}, \bibinfo {author} {\bibfnamefont {P.}~\bibnamefont {Quaino}},\
  and\ \bibinfo {author} {\bibfnamefont {W.}~\bibnamefont {Schmickler}},\
  }\href {https://doi.org/10.1039/C2CP40717E} {\bibfield  {journal} {\bibinfo
  {journal} {Phys. Chem. Chem. Phys.}\ }\textbf {\bibinfo {volume} {14}},\
  \bibinfo {pages} {11224} (\bibinfo {year} {2012})}\BibitemShut {NoStop}%
\bibitem [{\citenamefont {Hoffmann}(1989)}]{Hoffmann1988}%
  \BibitemOpen
  \bibfield  {author} {\bibinfo {author} {\bibfnamefont {R.}~\bibnamefont
  {Hoffmann}},\ }\href@noop {} {\emph {\bibinfo {title} {Solids and Surfaces, A
  Chemists View of Bonding in Extended Structures}}}\ (\bibinfo  {publisher}
  {Wiley-VCH Inc},\ \bibinfo {year} {1989})\BibitemShut {NoStop}%
\bibitem [{\citenamefont {Abild-Pedersen}\ \emph {et~al.}(2007)\citenamefont
  {Abild-Pedersen}, \citenamefont {Greeley}, \citenamefont {Studt},
  \citenamefont {Rossmeisl}, \citenamefont {Munter}, \citenamefont {Moses},
  \citenamefont {Sk\'ulason}, \citenamefont {Bligaard},\ and\ \citenamefont
  {N\o{}rskov}}]{Pedersen2007}%
  \BibitemOpen
  \bibfield  {author} {\bibinfo {author} {\bibfnamefont {F.}~\bibnamefont
  {Abild-Pedersen}}, \bibinfo {author} {\bibfnamefont {J.}~\bibnamefont
  {Greeley}}, \bibinfo {author} {\bibfnamefont {F.}~\bibnamefont {Studt}},
  \bibinfo {author} {\bibfnamefont {J.}~\bibnamefont {Rossmeisl}}, \bibinfo
  {author} {\bibfnamefont {T.~R.}\ \bibnamefont {Munter}}, \bibinfo {author}
  {\bibfnamefont {P.~G.}\ \bibnamefont {Moses}}, \bibinfo {author}
  {\bibfnamefont {E.}~\bibnamefont {Sk\'ulason}}, \bibinfo {author}
  {\bibfnamefont {T.}~\bibnamefont {Bligaard}},\ and\ \bibinfo {author}
  {\bibfnamefont {J.~K.}\ \bibnamefont {N\o{}rskov}},\ }\href
  {https://doi.org/10.1103/PhysRevLett.99.016105} {\bibfield  {journal}
  {\bibinfo  {journal} {Phys. Rev. Lett.}\ }\textbf {\bibinfo {volume} {99}},\
  \bibinfo {pages} {016105} (\bibinfo {year} {2007})}\BibitemShut {NoStop}%
\bibitem [{\citenamefont {Saini}\ \emph {et~al.}(2022)\citenamefont {Saini},
  \citenamefont {Halldin~Stenlid},\ and\ \citenamefont
  {Abild-Pedersen}}]{Pedersen2022}%
  \BibitemOpen
  \bibfield  {author} {\bibinfo {author} {\bibfnamefont {S.}~\bibnamefont
  {Saini}}, \bibinfo {author} {\bibfnamefont {J.}~\bibnamefont
  {Halldin~Stenlid}},\ and\ \bibinfo {author} {\bibfnamefont {F.}~\bibnamefont
  {Abild-Pedersen}},\ }\href {https://doi.org/10.1038/s41524-022-00846-z}
  {\bibfield  {journal} {\bibinfo  {journal} {npj Comput. Mater.}\ }\textbf
  {\bibinfo {volume} {8}},\ \bibinfo {pages} {163} (\bibinfo {year}
  {2022})}\BibitemShut {NoStop}%
\bibitem [{\citenamefont {Bligaard}\ and\ \citenamefont
  {Nørskov}(2008)}]{Bligaard2008}%
  \BibitemOpen
  \bibfield  {author} {\bibinfo {author} {\bibfnamefont {T.}~\bibnamefont
  {Bligaard}}\ and\ \bibinfo {author} {\bibfnamefont {J.}~\bibnamefont
  {Nørskov}},\ }in\ \href {https://doi.org/10.1016/B978-044452837-7.50005-8}
  {\emph {\bibinfo {booktitle} {Chemical Bonding at Surfaces and
  Interfaces}}},\ \bibinfo {editor} {edited by\ \bibinfo {editor}
  {\bibfnamefont {A.}~\bibnamefont {Nilsson}}, \bibinfo {editor} {\bibfnamefont
  {L.~G.}\ \bibnamefont {Pettersson}},\ and\ \bibinfo {editor} {\bibfnamefont
  {J.~K.}\ \bibnamefont {Nørskov}}}\ (\bibinfo  {publisher} {Elsevier},\
  \bibinfo {address} {Amsterdam},\ \bibinfo {year} {2008})\ pp.\ \bibinfo
  {pages} {255--321}\BibitemShut {NoStop}%
\bibitem [{\citenamefont {Nilsson}\ and\ \citenamefont
  {Pettersson}(2008)}]{Nilsson2008}%
  \BibitemOpen
  \bibfield  {author} {\bibinfo {author} {\bibfnamefont {A.}~\bibnamefont
  {Nilsson}}\ and\ \bibinfo {author} {\bibfnamefont {L.~G.}\ \bibnamefont
  {Pettersson}},\ }in\ \href {https://doi.org/10.1016/B978-044452837-7.50003-4}
  {\emph {\bibinfo {booktitle} {Chemical Bonding at Surfaces and
  Interfaces}}},\ \bibinfo {editor} {edited by\ \bibinfo {editor}
  {\bibfnamefont {A.}~\bibnamefont {Nilsson}}, \bibinfo {editor} {\bibfnamefont
  {L.~G.}\ \bibnamefont {Pettersson}},\ and\ \bibinfo {editor} {\bibfnamefont
  {J.~K.}\ \bibnamefont {Nørskov}}}\ (\bibinfo  {publisher} {Elsevier},\
  \bibinfo {address} {Amsterdam},\ \bibinfo {year} {2008})\ pp.\ \bibinfo
  {pages} {57--142}\BibitemShut {NoStop}%
\bibitem [{\citenamefont {Gross}(2003)}]{Gross2009}%
  \BibitemOpen
  \bibfield  {author} {\bibinfo {author} {\bibfnamefont {A.}~\bibnamefont
  {Gross}},\ }\href {https://doi.org/10.1007/978-3-662-05041-5} {\emph
  {\bibinfo {title} {Theoretical Surface Science, A Microscopic Perspective}}}\
  (\bibinfo  {publisher} {Springer},\ \bibinfo {year} {2003})\BibitemShut
  {NoStop}%
\bibitem [{\citenamefont {Cao}\ and\ \citenamefont
  {Nørskov}(2023)}]{Norskov2023}%
  \BibitemOpen
  \bibfield  {author} {\bibinfo {author} {\bibfnamefont {A.}~\bibnamefont
  {Cao}}\ and\ \bibinfo {author} {\bibfnamefont {J.~K.}\ \bibnamefont
  {Nørskov}},\ }\href {https://doi.org/10.1021/acscatal.2c06319} {\bibfield
  {journal} {\bibinfo  {journal} {ACS Catal.}\ }\textbf {\bibinfo {volume}
  {13}},\ \bibinfo {pages} {3456} (\bibinfo {year} {2023})}\BibitemShut
  {NoStop}%
\bibitem [{\citenamefont {Persson}\ and\ \citenamefont
  {Persson}(1980)}]{Persson1980}%
  \BibitemOpen
  \bibfield  {author} {\bibinfo {author} {\bibfnamefont {B.}~\bibnamefont
  {Persson}}\ and\ \bibinfo {author} {\bibfnamefont {M.}~\bibnamefont
  {Persson}},\ }\href
  {https://doi.org/https://doi.org/10.1016/0038-1098(80)90677-8} {\bibfield
  {journal} {\bibinfo  {journal} {Solid State Communications}\ }\textbf
  {\bibinfo {volume} {36}},\ \bibinfo {pages} {175} (\bibinfo {year}
  {1980})}\BibitemShut {NoStop}%
\bibitem [{\citenamefont {Saalfrank}(2006)}]{Saalfrank2006}%
  \BibitemOpen
  \bibfield  {author} {\bibinfo {author} {\bibfnamefont {P.}~\bibnamefont
  {Saalfrank}},\ }\href {https://doi.org/10.1021/cr0501691} {\bibfield
  {journal} {\bibinfo  {journal} {Chem. Rev.}\ }\textbf {\bibinfo {volume}
  {106}},\ \bibinfo {pages} {4116} (\bibinfo {year} {2006})}\BibitemShut
  {NoStop}%
\bibitem [{\citenamefont {Menzel}(2012)}]{Menzel2012}%
  \BibitemOpen
  \bibfield  {author} {\bibinfo {author} {\bibfnamefont {D.}~\bibnamefont
  {Menzel}},\ }\href {https://doi.org/10.1063/1.4746799} {\bibfield  {journal}
  {\bibinfo  {journal} {J. Chem. Phys.}\ }\textbf {\bibinfo {volume} {137}},\
  \bibinfo {pages} {091702} (\bibinfo {year} {2012})}\BibitemShut {NoStop}%
\bibitem [{\citenamefont {Menzel}\ and\ \citenamefont
  {Gomer}(1964)}]{Menzel1964}%
  \BibitemOpen
  \bibfield  {author} {\bibinfo {author} {\bibfnamefont {D.}~\bibnamefont
  {Menzel}}\ and\ \bibinfo {author} {\bibfnamefont {R.}~\bibnamefont {Gomer}},\
  }\href {https://doi.org/10.1063/1.1725730} {\bibfield  {journal} {\bibinfo
  {journal} {J. Chem. Phys.}\ }\textbf {\bibinfo {volume} {41}},\ \bibinfo
  {pages} {3311} (\bibinfo {year} {1964})}\BibitemShut {NoStop}%
\bibitem [{\citenamefont {Misewich}\ \emph {et~al.}(1992)\citenamefont
  {Misewich}, \citenamefont {Heinz},\ and\ \citenamefont
  {Newns}}]{Misewich1992}%
  \BibitemOpen
  \bibfield  {author} {\bibinfo {author} {\bibfnamefont {J.~A.}\ \bibnamefont
  {Misewich}}, \bibinfo {author} {\bibfnamefont {T.~F.}\ \bibnamefont
  {Heinz}},\ and\ \bibinfo {author} {\bibfnamefont {D.~M.}\ \bibnamefont
  {Newns}},\ }\href {https://doi.org/10.1103/PhysRevLett.68.3737} {\bibfield
  {journal} {\bibinfo  {journal} {Phys. Rev. Lett.}\ }\textbf {\bibinfo
  {volume} {68}},\ \bibinfo {pages} {3737} (\bibinfo {year}
  {1992})}\BibitemShut {NoStop}%
\bibitem [{\citenamefont {Bardeen}(1961)}]{Bardeen1961}%
  \BibitemOpen
  \bibfield  {author} {\bibinfo {author} {\bibfnamefont {J.}~\bibnamefont
  {Bardeen}},\ }\href {https://doi.org/10.1103/PhysRevLett.6.57} {\bibfield
  {journal} {\bibinfo  {journal} {Phys. Rev. Lett.}\ }\textbf {\bibinfo
  {volume} {6}},\ \bibinfo {pages} {57} (\bibinfo {year} {1961})}\BibitemShut
  {NoStop}%
\bibitem [{\citenamefont {Tersoff}\ and\ \citenamefont
  {Hamann}(1985)}]{Tersoff_Hamann_1984}%
  \BibitemOpen
  \bibfield  {author} {\bibinfo {author} {\bibfnamefont {J.}~\bibnamefont
  {Tersoff}}\ and\ \bibinfo {author} {\bibfnamefont {D.~R.}\ \bibnamefont
  {Hamann}},\ }\href {https://doi.org/10.1103/PhysRevB.31.805} {\bibfield
  {journal} {\bibinfo  {journal} {Phys. Rev. B}\ }\textbf {\bibinfo {volume}
  {31}},\ \bibinfo {pages} {805} (\bibinfo {year} {1985})}\BibitemShut
  {NoStop}%
\bibitem [{\citenamefont {Tersoff}\ and\ \citenamefont
  {Hamann}(1983)}]{TersoffHamann1983}%
  \BibitemOpen
  \bibfield  {author} {\bibinfo {author} {\bibfnamefont {J.}~\bibnamefont
  {Tersoff}}\ and\ \bibinfo {author} {\bibfnamefont {D.~R.}\ \bibnamefont
  {Hamann}},\ }\href {https://doi.org/10.1103/PhysRevLett.50.1998} {\bibfield
  {journal} {\bibinfo  {journal} {Phys. Rev. Lett.}\ }\textbf {\bibinfo
  {volume} {50}},\ \bibinfo {pages} {1998} (\bibinfo {year}
  {1983})}\BibitemShut {NoStop}%
\bibitem [{\citenamefont {Chen}(1990)}]{Chen1990}%
  \BibitemOpen
  \bibfield  {author} {\bibinfo {author} {\bibfnamefont {C.~J.}\ \bibnamefont
  {Chen}},\ }\href {https://doi.org/10.1103/PhysRevB.42.8841} {\bibfield
  {journal} {\bibinfo  {journal} {Phys. Rev. B}\ }\textbf {\bibinfo {volume}
  {42}},\ \bibinfo {pages} {8841} (\bibinfo {year} {1990})}\BibitemShut
  {NoStop}%
\bibitem [{\citenamefont {Chen}(1992)}]{Chen1992}%
  \BibitemOpen
  \bibfield  {author} {\bibinfo {author} {\bibfnamefont {C.~J.}\ \bibnamefont
  {Chen}},\ }\href {https://doi.org/10.1103/PhysRevLett.69.1656} {\bibfield
  {journal} {\bibinfo  {journal} {Phys. Rev. Lett.}\ }\textbf {\bibinfo
  {volume} {69}},\ \bibinfo {pages} {1656} (\bibinfo {year}
  {1992})}\BibitemShut {NoStop}%
\bibitem [{\citenamefont {Chen}(2007)}]{Chen2007}%
  \BibitemOpen
  \bibfield  {author} {\bibinfo {author} {\bibfnamefont {C.~J.}\ \bibnamefont
  {Chen}},\ }\href {https://doi.org/10.1093/acprof:oso/9780199211500.001.0001}
  {\emph {\bibinfo {title} {{Introduction to Scanning Tunneling Microscopy}}}}\
  (\bibinfo  {publisher} {Oxford University Press},\ \bibinfo {year}
  {2007})\BibitemShut {NoStop}%
\bibitem [{\citenamefont {Cederbaum}\ \emph {et~al.}(1981)\citenamefont
  {Cederbaum}, \citenamefont {K\"oppel},\ and\ \citenamefont
  {Domcke}}]{Domcke1981_original_BD_paper}%
  \BibitemOpen
  \bibfield  {author} {\bibinfo {author} {\bibfnamefont {L.~S.}\ \bibnamefont
  {Cederbaum}}, \bibinfo {author} {\bibfnamefont {H.}~\bibnamefont
  {K\"oppel}},\ and\ \bibinfo {author} {\bibfnamefont {W.}~\bibnamefont
  {Domcke}},\ }\href {https://doi.org/10.1002/qua.560200828} {\bibfield
  {journal} {\bibinfo  {journal} {Int. J. Quantum Chem.}\ }\textbf {\bibinfo
  {volume} {20}},\ \bibinfo {pages} {251} (\bibinfo {year} {1981})}\BibitemShut
  {NoStop}%
\bibitem [{\citenamefont {Oberhofer}\ \emph {et~al.}(2017)\citenamefont
  {Oberhofer}, \citenamefont {Reuter},\ and\ \citenamefont
  {Blumberger}}]{Oberhofer2017}%
  \BibitemOpen
  \bibfield  {author} {\bibinfo {author} {\bibfnamefont {H.}~\bibnamefont
  {Oberhofer}}, \bibinfo {author} {\bibfnamefont {K.}~\bibnamefont {Reuter}},\
  and\ \bibinfo {author} {\bibfnamefont {J.}~\bibnamefont {Blumberger}},\
  }\href {https://doi.org/10.1021/acs.chemrev.7b00086} {\bibfield  {journal}
  {\bibinfo  {journal} {Chem. Rev.}\ }\textbf {\bibinfo {volume} {117}},\
  \bibinfo {pages} {10319} (\bibinfo {year} {2017})}\BibitemShut {NoStop}%
\bibitem [{\citenamefont {Moon}\ \emph {et~al.}(2008)\citenamefont {Moon},
  \citenamefont {Mattos}, \citenamefont {Foster}, \citenamefont {Zeltzer},
  \citenamefont {Ko},\ and\ \citenamefont {Manoharan}}]{Moon2008}%
  \BibitemOpen
  \bibfield  {author} {\bibinfo {author} {\bibfnamefont {C.~R.}\ \bibnamefont
  {Moon}}, \bibinfo {author} {\bibfnamefont {L.~S.}\ \bibnamefont {Mattos}},
  \bibinfo {author} {\bibfnamefont {B.~K.}\ \bibnamefont {Foster}}, \bibinfo
  {author} {\bibfnamefont {G.}~\bibnamefont {Zeltzer}}, \bibinfo {author}
  {\bibfnamefont {W.}~\bibnamefont {Ko}},\ and\ \bibinfo {author}
  {\bibfnamefont {H.~C.}\ \bibnamefont {Manoharan}},\ }\href
  {https://doi.org/10.1126/science.1151490} {\bibfield  {journal} {\bibinfo
  {journal} {Science}\ }\textbf {\bibinfo {volume} {319}},\ \bibinfo {pages}
  {782} (\bibinfo {year} {2008})}\BibitemShut {NoStop}%
\bibitem [{\citenamefont {Van~Voorhis}\ \emph {et~al.}(2010)\citenamefont
  {Van~Voorhis}, \citenamefont {Kowalczyk}, \citenamefont {Kaduk},
  \citenamefont {Wang}, \citenamefont {Cheng},\ and\ \citenamefont
  {Wu}}]{Voorhis2010}%
  \BibitemOpen
  \bibfield  {author} {\bibinfo {author} {\bibfnamefont {T.}~\bibnamefont
  {Van~Voorhis}}, \bibinfo {author} {\bibfnamefont {T.}~\bibnamefont
  {Kowalczyk}}, \bibinfo {author} {\bibfnamefont {B.}~\bibnamefont {Kaduk}},
  \bibinfo {author} {\bibfnamefont {L.-P.}\ \bibnamefont {Wang}}, \bibinfo
  {author} {\bibfnamefont {C.-L.}\ \bibnamefont {Cheng}},\ and\ \bibinfo
  {author} {\bibfnamefont {Q.}~\bibnamefont {Wu}},\ }\href
  {https://doi.org/10.1146/annurev.physchem.012809.103324} {\bibfield
  {journal} {\bibinfo  {journal} {Annu. Rev. Phys. Chem.}\ }\textbf {\bibinfo
  {volume} {61}},\ \bibinfo {pages} {149} (\bibinfo {year} {2010})}\BibitemShut
  {NoStop}%
\bibitem [{\citenamefont {Ghan}\ \emph {et~al.}(2020)\citenamefont {Ghan},
  \citenamefont {Kunkel}, \citenamefont {Reuter},\ and\ \citenamefont
  {Oberhofer}}]{Ghan2020}%
  \BibitemOpen
  \bibfield  {author} {\bibinfo {author} {\bibfnamefont {S.}~\bibnamefont
  {Ghan}}, \bibinfo {author} {\bibfnamefont {C.}~\bibnamefont {Kunkel}},
  \bibinfo {author} {\bibfnamefont {K.}~\bibnamefont {Reuter}},\ and\ \bibinfo
  {author} {\bibfnamefont {H.}~\bibnamefont {Oberhofer}},\ }\href
  {https://doi.org/10.1021/acs.jctc.0c00887} {\bibfield  {journal} {\bibinfo
  {journal} {J. Chem. Theory Comput.}\ }\textbf {\bibinfo {volume} {16}},\
  \bibinfo {pages} {7431} (\bibinfo {year} {2020})}\BibitemShut {NoStop}%
\bibitem [{\citenamefont {Liu}\ \emph {et~al.}(2021)\citenamefont {Liu},
  \citenamefont {Kang}, \citenamefont {Perry}, \citenamefont {Chen},
  \citenamefont {West}, \citenamefont {Xia}, \citenamefont {Chaudhuri},
  \citenamefont {Laker}, \citenamefont {Wilson}, \citenamefont {Meloni},
  \citenamefont {Melander}, \citenamefont {Maurer},\ and\ \citenamefont
  {Unwin}}]{Maurer2021}%
  \BibitemOpen
  \bibfield  {author} {\bibinfo {author} {\bibfnamefont {D.-Q.}\ \bibnamefont
  {Liu}}, \bibinfo {author} {\bibfnamefont {M.}~\bibnamefont {Kang}}, \bibinfo
  {author} {\bibfnamefont {D.}~\bibnamefont {Perry}}, \bibinfo {author}
  {\bibfnamefont {C.-H.}\ \bibnamefont {Chen}}, \bibinfo {author}
  {\bibfnamefont {G.}~\bibnamefont {West}}, \bibinfo {author} {\bibfnamefont
  {X.}~\bibnamefont {Xia}}, \bibinfo {author} {\bibfnamefont {S.}~\bibnamefont
  {Chaudhuri}}, \bibinfo {author} {\bibfnamefont {Z.~P.~L.}\ \bibnamefont
  {Laker}}, \bibinfo {author} {\bibfnamefont {N.~R.}\ \bibnamefont {Wilson}},
  \bibinfo {author} {\bibfnamefont {G.~N.}\ \bibnamefont {Meloni}}, \bibinfo
  {author} {\bibfnamefont {M.~M.}\ \bibnamefont {Melander}}, \bibinfo {author}
  {\bibfnamefont {R.~J.}\ \bibnamefont {Maurer}},\ and\ \bibinfo {author}
  {\bibfnamefont {P.~R.}\ \bibnamefont {Unwin}},\ }\href
  {https://doi.org/10.1038/s41467-021-27339-9} {\bibfield  {journal} {\bibinfo
  {journal} {Nat. Commun.}\ }\textbf {\bibinfo {volume} {12}},\ \bibinfo
  {pages} {7110} (\bibinfo {year} {2021})}\BibitemShut {NoStop}%
\bibitem [{\citenamefont {Vojvodic}\ \emph {et~al.}(2014)\citenamefont
  {Vojvodic}, \citenamefont {N{\o}rskov},\ and\ \citenamefont
  {Abild-Pedersen}}]{Pedersen2014}%
  \BibitemOpen
  \bibfield  {author} {\bibinfo {author} {\bibfnamefont {A.}~\bibnamefont
  {Vojvodic}}, \bibinfo {author} {\bibfnamefont {J.~K.}\ \bibnamefont
  {N{\o}rskov}},\ and\ \bibinfo {author} {\bibfnamefont {F.}~\bibnamefont
  {Abild-Pedersen}},\ }\href {https://doi.org/10.1007/s11244-013-0159-2}
  {\bibfield  {journal} {\bibinfo  {journal} {Top. Catal.}\ }\textbf {\bibinfo
  {volume} {57}},\ \bibinfo {pages} {25} (\bibinfo {year} {2014})}\BibitemShut
  {NoStop}%
\bibitem [{\citenamefont {Keller}\ \emph
  {et~al.}(1998{\natexlab{a}})\citenamefont {Keller}, \citenamefont {Stichler},
  \citenamefont {Comelli}, \citenamefont {Esch}, \citenamefont {Lizzit},
  \citenamefont {Menzel},\ and\ \citenamefont {Wurth}}]{Menzel_1998}%
  \BibitemOpen
  \bibfield  {author} {\bibinfo {author} {\bibfnamefont {C.}~\bibnamefont
  {Keller}}, \bibinfo {author} {\bibfnamefont {M.}~\bibnamefont {Stichler}},
  \bibinfo {author} {\bibfnamefont {G.}~\bibnamefont {Comelli}}, \bibinfo
  {author} {\bibfnamefont {F.}~\bibnamefont {Esch}}, \bibinfo {author}
  {\bibfnamefont {S.}~\bibnamefont {Lizzit}}, \bibinfo {author} {\bibfnamefont
  {D.}~\bibnamefont {Menzel}},\ and\ \bibinfo {author} {\bibfnamefont
  {W.}~\bibnamefont {Wurth}},\ }\href
  {https://doi.org/10.1103/PhysRevB.57.11951} {\bibfield  {journal} {\bibinfo
  {journal} {Phys. Rev. B}\ }\textbf {\bibinfo {volume} {57}},\ \bibinfo
  {pages} {11951} (\bibinfo {year} {1998}{\natexlab{a}})}\BibitemShut {NoStop}%
\bibitem [{\citenamefont {Menzel}(2008)}]{Menzel_2008}%
  \BibitemOpen
  \bibfield  {author} {\bibinfo {author} {\bibfnamefont {D.}~\bibnamefont
  {Menzel}},\ }\href {https://doi.org/10.1039/B719546J} {\bibfield  {journal}
  {\bibinfo  {journal} {Chem. Soc. Rev.}\ }\textbf {\bibinfo {volume} {37}},\
  \bibinfo {pages} {2212} (\bibinfo {year} {2008})}\BibitemShut {NoStop}%
\bibitem [{\citenamefont {Lizzit}\ \emph {et~al.}(2009)\citenamefont {Lizzit},
  \citenamefont {Zampieri}, \citenamefont {Kostov}, \citenamefont {Tyuliev},
  \citenamefont {Larciprete}, \citenamefont {Petaccia}, \citenamefont
  {Naydenov},\ and\ \citenamefont {Menzel}}]{Menzel_2009}%
  \BibitemOpen
  \bibfield  {author} {\bibinfo {author} {\bibfnamefont {S.}~\bibnamefont
  {Lizzit}}, \bibinfo {author} {\bibfnamefont {G.}~\bibnamefont {Zampieri}},
  \bibinfo {author} {\bibfnamefont {K.~L.}\ \bibnamefont {Kostov}}, \bibinfo
  {author} {\bibfnamefont {G.}~\bibnamefont {Tyuliev}}, \bibinfo {author}
  {\bibfnamefont {R.}~\bibnamefont {Larciprete}}, \bibinfo {author}
  {\bibfnamefont {L.}~\bibnamefont {Petaccia}}, \bibinfo {author}
  {\bibfnamefont {B.}~\bibnamefont {Naydenov}},\ and\ \bibinfo {author}
  {\bibfnamefont {D.}~\bibnamefont {Menzel}},\ }\href
  {https://doi.org/10.1088/1367-2630/11/5/053005} {\bibfield  {journal}
  {\bibinfo  {journal} {New J. Phys.}\ }\textbf {\bibinfo {volume} {11}},\
  \bibinfo {pages} {053005} (\bibinfo {year} {2009})}\BibitemShut {NoStop}%
\bibitem [{\citenamefont {Schlichting}\ \emph {et~al.}(1988)\citenamefont
  {Schlichting}, \citenamefont {Menzel}, \citenamefont {Brunner}, \citenamefont
  {Brenig},\ and\ \citenamefont {Tully}}]{Menzel1988}%
  \BibitemOpen
  \bibfield  {author} {\bibinfo {author} {\bibfnamefont {H.}~\bibnamefont
  {Schlichting}}, \bibinfo {author} {\bibfnamefont {D.}~\bibnamefont {Menzel}},
  \bibinfo {author} {\bibfnamefont {T.}~\bibnamefont {Brunner}}, \bibinfo
  {author} {\bibfnamefont {W.}~\bibnamefont {Brenig}},\ and\ \bibinfo {author}
  {\bibfnamefont {J.~C.}\ \bibnamefont {Tully}},\ }\href
  {https://doi.org/10.1103/PhysRevLett.60.2515} {\bibfield  {journal} {\bibinfo
   {journal} {Phys. Rev. Lett.}\ }\textbf {\bibinfo {volume} {60}},\ \bibinfo
  {pages} {2515} (\bibinfo {year} {1988})}\BibitemShut {NoStop}%
\bibitem [{\citenamefont {Föhlisch}\ \emph {et~al.}(2003)\citenamefont
  {Föhlisch}, \citenamefont {Menzel}, \citenamefont {Feulner}, \citenamefont
  {Ecker}, \citenamefont {Weimar}, \citenamefont {Kostov}, \citenamefont
  {Tyuliev}, \citenamefont {Lizzit}, \citenamefont {Larciprete}, \citenamefont
  {Hennies},\ and\ \citenamefont {Wurth}}]{Wurth2002}%
  \BibitemOpen
  \bibfield  {author} {\bibinfo {author} {\bibfnamefont {A.}~\bibnamefont
  {Föhlisch}}, \bibinfo {author} {\bibfnamefont {D.}~\bibnamefont {Menzel}},
  \bibinfo {author} {\bibfnamefont {P.}~\bibnamefont {Feulner}}, \bibinfo
  {author} {\bibfnamefont {M.}~\bibnamefont {Ecker}}, \bibinfo {author}
  {\bibfnamefont {R.}~\bibnamefont {Weimar}}, \bibinfo {author} {\bibfnamefont
  {K.}~\bibnamefont {Kostov}}, \bibinfo {author} {\bibfnamefont
  {G.}~\bibnamefont {Tyuliev}}, \bibinfo {author} {\bibfnamefont
  {S.}~\bibnamefont {Lizzit}}, \bibinfo {author} {\bibfnamefont
  {R.}~\bibnamefont {Larciprete}}, \bibinfo {author} {\bibfnamefont
  {F.}~\bibnamefont {Hennies}},\ and\ \bibinfo {author} {\bibfnamefont
  {W.}~\bibnamefont {Wurth}},\ }\href
  {https://doi.org/10.1016/S0301-0104(02)00939-4} {\bibfield  {journal}
  {\bibinfo  {journal} {Chem. Phys.}\ }\textbf {\bibinfo {volume} {289}},\
  \bibinfo {pages} {107} (\bibinfo {year} {2003})}\BibitemShut {NoStop}%
\bibitem [{\citenamefont {Föhlisch}\ \emph {et~al.}(2005)\citenamefont
  {Föhlisch}, \citenamefont {Feulner}, \citenamefont {Hennies}, \citenamefont
  {Fink}, \citenamefont {Menzel}, \citenamefont {S\'anchez-Portal},
  \citenamefont {M~Echenique},\ and\ \citenamefont {Wurth}}]{Feulner_2005}%
  \BibitemOpen
  \bibfield  {author} {\bibinfo {author} {\bibfnamefont {A.}~\bibnamefont
  {Föhlisch}}, \bibinfo {author} {\bibfnamefont {P.}~\bibnamefont {Feulner}},
  \bibinfo {author} {\bibfnamefont {F.}~\bibnamefont {Hennies}}, \bibinfo
  {author} {\bibfnamefont {A.}~\bibnamefont {Fink}}, \bibinfo {author}
  {\bibfnamefont {D.}~\bibnamefont {Menzel}}, \bibinfo {author} {\bibfnamefont
  {D.}~\bibnamefont {S\'anchez-Portal}}, \bibinfo {author} {\bibfnamefont
  {P.}~\bibnamefont {M~Echenique}},\ and\ \bibinfo {author} {\bibfnamefont
  {W.}~\bibnamefont {Wurth}},\ }\href {https://doi.org/10.1038/nature03833}
  {\bibfield  {journal} {\bibinfo  {journal} {Nature}\ }\textbf {\bibinfo
  {volume} {436}},\ \bibinfo {pages} {373} (\bibinfo {year}
  {2005})}\BibitemShut {NoStop}%
\bibitem [{\citenamefont {Neppl}\ \emph {et~al.}(2007)\citenamefont {Neppl},
  \citenamefont {Bauer}, \citenamefont {Menzel}, \citenamefont {Feulner},
  \citenamefont {Shaporenko}, \citenamefont {Zharnikov}, \citenamefont {Kao},\
  and\ \citenamefont {Allara}}]{Feulner_2007}%
  \BibitemOpen
  \bibfield  {author} {\bibinfo {author} {\bibfnamefont {S.}~\bibnamefont
  {Neppl}}, \bibinfo {author} {\bibfnamefont {U.}~\bibnamefont {Bauer}},
  \bibinfo {author} {\bibfnamefont {D.}~\bibnamefont {Menzel}}, \bibinfo
  {author} {\bibfnamefont {P.}~\bibnamefont {Feulner}}, \bibinfo {author}
  {\bibfnamefont {A.}~\bibnamefont {Shaporenko}}, \bibinfo {author}
  {\bibfnamefont {M.}~\bibnamefont {Zharnikov}}, \bibinfo {author}
  {\bibfnamefont {P.}~\bibnamefont {Kao}},\ and\ \bibinfo {author}
  {\bibfnamefont {D.}~\bibnamefont {Allara}},\ }\href
  {https://doi.org/10.1016/j.cplett.2007.09.013} {\bibfield  {journal}
  {\bibinfo  {journal} {Chem. Phys. Lett.}\ }\textbf {\bibinfo {volume}
  {447}},\ \bibinfo {pages} {227 } (\bibinfo {year} {2007})}\BibitemShut
  {NoStop}%
\bibitem [{\citenamefont {Kao}\ \emph {et~al.}(2010)\citenamefont {Kao},
  \citenamefont {Neppl}, \citenamefont {Feulner}, \citenamefont {Allara},\ and\
  \citenamefont {Zharnikov}}]{Feulner_2010}%
  \BibitemOpen
  \bibfield  {author} {\bibinfo {author} {\bibfnamefont {P.}~\bibnamefont
  {Kao}}, \bibinfo {author} {\bibfnamefont {S.}~\bibnamefont {Neppl}}, \bibinfo
  {author} {\bibfnamefont {P.}~\bibnamefont {Feulner}}, \bibinfo {author}
  {\bibfnamefont {D.~L.}\ \bibnamefont {Allara}},\ and\ \bibinfo {author}
  {\bibfnamefont {M.}~\bibnamefont {Zharnikov}},\ }\href
  {https://doi.org/10.1021/jp1042816} {\bibfield  {journal} {\bibinfo
  {journal} {J. Phys. Chem. C}\ }\textbf {\bibinfo {volume} {114}},\ \bibinfo
  {pages} {13766} (\bibinfo {year} {2010})}\BibitemShut {NoStop}%
\bibitem [{\citenamefont {Hamoudi}\ \emph {et~al.}(2011)\citenamefont
  {Hamoudi}, \citenamefont {Neppl}, \citenamefont {Kao}, \citenamefont
  {Sch\"upbach}, \citenamefont {Feulner}, \citenamefont {Terfort},
  \citenamefont {Allara},\ and\ \citenamefont {Zharnikov}}]{Feulner_2011}%
  \BibitemOpen
  \bibfield  {author} {\bibinfo {author} {\bibfnamefont {H.}~\bibnamefont
  {Hamoudi}}, \bibinfo {author} {\bibfnamefont {S.}~\bibnamefont {Neppl}},
  \bibinfo {author} {\bibfnamefont {P.}~\bibnamefont {Kao}}, \bibinfo {author}
  {\bibfnamefont {B.}~\bibnamefont {Sch\"upbach}}, \bibinfo {author}
  {\bibfnamefont {P.}~\bibnamefont {Feulner}}, \bibinfo {author} {\bibfnamefont
  {A.}~\bibnamefont {Terfort}}, \bibinfo {author} {\bibfnamefont
  {D.}~\bibnamefont {Allara}},\ and\ \bibinfo {author} {\bibfnamefont
  {M.}~\bibnamefont {Zharnikov}},\ }\href
  {https://doi.org/10.1103/PhysRevLett.107.027801} {\bibfield  {journal}
  {\bibinfo  {journal} {Phys. Rev. Lett.}\ }\textbf {\bibinfo {volume} {107}},\
  \bibinfo {pages} {027801} (\bibinfo {year} {2011})}\BibitemShut {NoStop}%
\bibitem [{\citenamefont {Blobner}\ \emph {et~al.}(2014)\citenamefont
  {Blobner}, \citenamefont {Han}, \citenamefont {Kim}, \citenamefont {Wurth},\
  and\ \citenamefont {Feulner}}]{Feulner_2014}%
  \BibitemOpen
  \bibfield  {author} {\bibinfo {author} {\bibfnamefont {F.}~\bibnamefont
  {Blobner}}, \bibinfo {author} {\bibfnamefont {R.}~\bibnamefont {Han}},
  \bibinfo {author} {\bibfnamefont {A.}~\bibnamefont {Kim}}, \bibinfo {author}
  {\bibfnamefont {W.}~\bibnamefont {Wurth}},\ and\ \bibinfo {author}
  {\bibfnamefont {P.}~\bibnamefont {Feulner}},\ }\href
  {https://doi.org/10.1103/PhysRevLett.112.086801} {\bibfield  {journal}
  {\bibinfo  {journal} {Phys. Rev. Lett.}\ }\textbf {\bibinfo {volume} {112}},\
  \bibinfo {pages} {086801} (\bibinfo {year} {2014})}\BibitemShut {NoStop}%
\bibitem [{\citenamefont {Feulner}\ \emph {et~al.}(2015)\citenamefont
  {Feulner}, \citenamefont {Blobner}, \citenamefont {Bauer}, \citenamefont
  {Han}, \citenamefont {Kim}, \citenamefont {Sundermann}, \citenamefont
  {Mueller}, \citenamefont {Heinzmann},\ and\ \citenamefont
  {Wurth}}]{Feulner_2015}%
  \BibitemOpen
  \bibfield  {author} {\bibinfo {author} {\bibfnamefont {P.}~\bibnamefont
  {Feulner}}, \bibinfo {author} {\bibfnamefont {F.}~\bibnamefont {Blobner}},
  \bibinfo {author} {\bibfnamefont {J.}~\bibnamefont {Bauer}}, \bibinfo
  {author} {\bibfnamefont {R.}~\bibnamefont {Han}}, \bibinfo {author}
  {\bibfnamefont {A.}~\bibnamefont {Kim}}, \bibinfo {author} {\bibfnamefont
  {T.}~\bibnamefont {Sundermann}}, \bibinfo {author} {\bibfnamefont
  {N.}~\bibnamefont {Mueller}}, \bibinfo {author} {\bibfnamefont
  {U.}~\bibnamefont {Heinzmann}},\ and\ \bibinfo {author} {\bibfnamefont
  {W.}~\bibnamefont {Wurth}},\ }\href {https://doi.org/10.1380/ejssnt.2015.317}
  {\bibfield  {journal} {\bibinfo  {journal} {e-J. Surf. Sci. Nanotechnol.}\
  }\textbf {\bibinfo {volume} {13}},\ \bibinfo {pages} {317} (\bibinfo {year}
  {2015})}\BibitemShut {NoStop}%
\bibitem [{\citenamefont {Sundermann}\ \emph {et~al.}(2016)\citenamefont
  {Sundermann}, \citenamefont {Müller}, \citenamefont {Heinzmann},
  \citenamefont {Wurth}, \citenamefont {Bauer}, \citenamefont {Han},
  \citenamefont {Kim}, \citenamefont {Menzel},\ and\ \citenamefont
  {Feulner}}]{Feulner2016}%
  \BibitemOpen
  \bibfield  {author} {\bibinfo {author} {\bibfnamefont {T.}~\bibnamefont
  {Sundermann}}, \bibinfo {author} {\bibfnamefont {N.}~\bibnamefont {Müller}},
  \bibinfo {author} {\bibfnamefont {U.}~\bibnamefont {Heinzmann}}, \bibinfo
  {author} {\bibfnamefont {W.}~\bibnamefont {Wurth}}, \bibinfo {author}
  {\bibfnamefont {J.}~\bibnamefont {Bauer}}, \bibinfo {author} {\bibfnamefont
  {R.}~\bibnamefont {Han}}, \bibinfo {author} {\bibfnamefont {A.}~\bibnamefont
  {Kim}}, \bibinfo {author} {\bibfnamefont {D.}~\bibnamefont {Menzel}},\ and\
  \bibinfo {author} {\bibfnamefont {P.}~\bibnamefont {Feulner}},\ }\href
  {https://doi.org/10.1016/j.susc.2015.08.031} {\bibfield  {journal} {\bibinfo
  {journal} {Surf. Sci.}\ }\textbf {\bibinfo {volume} {643}},\ \bibinfo {pages}
  {190 } (\bibinfo {year} {2016})}\BibitemShut {NoStop}%
\bibitem [{\citenamefont {Gauyacq}\ \emph {et~al.}(2001)\citenamefont
  {Gauyacq}, \citenamefont {Borisov},\ and\ \citenamefont
  {Raseev}}]{Raseev2001}%
  \BibitemOpen
  \bibfield  {author} {\bibinfo {author} {\bibfnamefont {J.}~\bibnamefont
  {Gauyacq}}, \bibinfo {author} {\bibfnamefont {A.}~\bibnamefont {Borisov}},\
  and\ \bibinfo {author} {\bibfnamefont {G.}~\bibnamefont {Raseev}},\ }\href
  {https://doi.org/10.1016/S0039-6028(01)01229-8} {\bibfield  {journal}
  {\bibinfo  {journal} {Surf. Sci.}\ }\textbf {\bibinfo {volume} {490}},\
  \bibinfo {pages} {99} (\bibinfo {year} {2001})}\BibitemShut {NoStop}%
\bibitem [{\citenamefont {S\'anchez-Portal}\ \emph {et~al.}(2007)\citenamefont
  {S\'anchez-Portal}, \citenamefont {Menzel},\ and\ \citenamefont
  {Echenique}}]{Echenique2007}%
  \BibitemOpen
  \bibfield  {author} {\bibinfo {author} {\bibfnamefont {D.}~\bibnamefont
  {S\'anchez-Portal}}, \bibinfo {author} {\bibfnamefont {D.}~\bibnamefont
  {Menzel}},\ and\ \bibinfo {author} {\bibfnamefont {P.~M.}\ \bibnamefont
  {Echenique}},\ }\href {https://doi.org/10.1103/PhysRevB.76.235406} {\bibfield
   {journal} {\bibinfo  {journal} {Phys. Rev. B}\ }\textbf {\bibinfo {volume}
  {76}},\ \bibinfo {pages} {235406} (\bibinfo {year} {2007})}\BibitemShut
  {NoStop}%
\bibitem [{\citenamefont {Wurth}\ and\ \citenamefont
  {Menzel}(2000)}]{Menzel2000}%
  \BibitemOpen
  \bibfield  {author} {\bibinfo {author} {\bibfnamefont {W.}~\bibnamefont
  {Wurth}}\ and\ \bibinfo {author} {\bibfnamefont {D.}~\bibnamefont {Menzel}},\
  }\href {https://doi.org/10.1016/S0301-0104(99)00305-5} {\bibfield  {journal}
  {\bibinfo  {journal} {Chem. Phys.}\ }\textbf {\bibinfo {volume} {251}},\
  \bibinfo {pages} {141} (\bibinfo {year} {2000})}\BibitemShut {NoStop}%
\bibitem [{\citenamefont {Vijayalakshmi}\ \emph {et~al.}(2006)\citenamefont
  {Vijayalakshmi}, \citenamefont {F\"ohlisch}, \citenamefont {Hennies},
  \citenamefont {Pietzsch}, \citenamefont {Nagasono}, \citenamefont {Wurth},
  \citenamefont {Borisov},\ and\ \citenamefont {Gauyacq}}]{Gauyacq2006}%
  \BibitemOpen
  \bibfield  {author} {\bibinfo {author} {\bibfnamefont {S.}~\bibnamefont
  {Vijayalakshmi}}, \bibinfo {author} {\bibfnamefont {A.}~\bibnamefont
  {F\"ohlisch}}, \bibinfo {author} {\bibfnamefont {F.}~\bibnamefont {Hennies}},
  \bibinfo {author} {\bibfnamefont {A.}~\bibnamefont {Pietzsch}}, \bibinfo
  {author} {\bibfnamefont {M.}~\bibnamefont {Nagasono}}, \bibinfo {author}
  {\bibfnamefont {W.}~\bibnamefont {Wurth}}, \bibinfo {author} {\bibfnamefont
  {A.}~\bibnamefont {Borisov}},\ and\ \bibinfo {author} {\bibfnamefont
  {J.}~\bibnamefont {Gauyacq}},\ }\href
  {https://doi.org/10.1016/j.cplett.2006.06.062} {\bibfield  {journal}
  {\bibinfo  {journal} {Chem. Phys. Lett.}\ }\textbf {\bibinfo {volume}
  {427}},\ \bibinfo {pages} {91 } (\bibinfo {year} {2006})}\BibitemShut
  {NoStop}%
\bibitem [{\citenamefont {Gauyacq}\ and\ \citenamefont
  {Borisov}(2004)}]{Borisov2004}%
  \BibitemOpen
  \bibfield  {author} {\bibinfo {author} {\bibfnamefont {J.~P.}\ \bibnamefont
  {Gauyacq}}\ and\ \bibinfo {author} {\bibfnamefont {A.~G.}\ \bibnamefont
  {Borisov}},\ }\href {https://doi.org/10.1103/PhysRevB.69.235408} {\bibfield
  {journal} {\bibinfo  {journal} {Phys. Rev. B}\ }\textbf {\bibinfo {volume}
  {69}},\ \bibinfo {pages} {235408} (\bibinfo {year} {2004})}\BibitemShut
  {NoStop}%
\bibitem [{\citenamefont {Müller}\ \emph {et~al.}(2020)\citenamefont
  {Müller}, \citenamefont {Echenique},\ and\ \citenamefont
  {S\'anchez-Portal}}]{Muller2020}%
  \BibitemOpen
  \bibfield  {author} {\bibinfo {author} {\bibfnamefont {M.}~\bibnamefont
  {Müller}}, \bibinfo {author} {\bibfnamefont {P.~M.}\ \bibnamefont
  {Echenique}},\ and\ \bibinfo {author} {\bibfnamefont {D.}~\bibnamefont
  {S\'anchez-Portal}},\ }\href {https://doi.org/10.1021/acs.jpclett.0c01946}
  {\bibfield  {journal} {\bibinfo  {journal} {J. Phys. Chem. Lett.}\ }\textbf
  {\bibinfo {volume} {11}},\ \bibinfo {pages} {7141} (\bibinfo {year}
  {2020})}\BibitemShut {NoStop}%
\bibitem [{\citenamefont {Newns}(1969)}]{Newns1969}%
  \BibitemOpen
  \bibfield  {author} {\bibinfo {author} {\bibfnamefont {D.}~\bibnamefont
  {Newns}},\ }\href {https://doi.org/10.1103/PhysRev.178.1123} {\bibfield
  {journal} {\bibinfo  {journal} {Phys. Rev.}\ }\textbf {\bibinfo {volume}
  {178}},\ \bibinfo {pages} {1123} (\bibinfo {year} {1969})}\BibitemShut
  {NoStop}%
\bibitem [{\citenamefont {Anderson}(1961)}]{Anderson1961}%
  \BibitemOpen
  \bibfield  {author} {\bibinfo {author} {\bibfnamefont {P.~W.}\ \bibnamefont
  {Anderson}},\ }\href {https://doi.org/10.1103/PhysRev.124.41} {\bibfield
  {journal} {\bibinfo  {journal} {Phys. Rev.}\ }\textbf {\bibinfo {volume}
  {124}},\ \bibinfo {pages} {41} (\bibinfo {year} {1961})}\BibitemShut
  {NoStop}%
\bibitem [{\citenamefont {Grimley}(1967)}]{Grimley1967}%
  \BibitemOpen
  \bibfield  {author} {\bibinfo {author} {\bibfnamefont {T.~B.}\ \bibnamefont
  {Grimley}},\ }\href {https://doi.org/10.1088/0370-1328/90/3/320} {\bibfield
  {journal} {\bibinfo  {journal} {Proc. Phys. Soc.}\ }\textbf {\bibinfo
  {volume} {90}},\ \bibinfo {pages} {751} (\bibinfo {year} {1967})}\BibitemShut
  {NoStop}%
\bibitem [{\citenamefont {Kondov}\ \emph {et~al.}(2007)\citenamefont {Kondov},
  \citenamefont {\v{C}\'{\i}\v{z}ek}, \citenamefont {Benesch}, \citenamefont
  {Wang},\ and\ \citenamefont {Thoss}}]{Kondov2007}%
  \BibitemOpen
  \bibfield  {author} {\bibinfo {author} {\bibfnamefont {I.}~\bibnamefont
  {Kondov}}, \bibinfo {author} {\bibfnamefont {M.}~\bibnamefont
  {\v{C}\'{\i}\v{z}ek}}, \bibinfo {author} {\bibfnamefont {C.}~\bibnamefont
  {Benesch}}, \bibinfo {author} {\bibfnamefont {H.}~\bibnamefont {Wang}},\ and\
  \bibinfo {author} {\bibfnamefont {M.}~\bibnamefont {Thoss}},\ }\href
  {https://doi.org/10.1021/jp072217m} {\bibfield  {journal} {\bibinfo
  {journal} {J. Phys. Chem. C}\ }\textbf {\bibinfo {volume} {111}},\ \bibinfo
  {pages} {11970} (\bibinfo {year} {2007})}\BibitemShut {NoStop}%
\bibitem [{\citenamefont {Li}\ \emph {et~al.}(2008)\citenamefont {Li},
  \citenamefont {Nilsing}, \citenamefont {Kondov}, \citenamefont {Wang},
  \citenamefont {Persson}, \citenamefont {Lunell},\ and\ \citenamefont
  {Thoss}}]{Jingrui2008}%
  \BibitemOpen
  \bibfield  {author} {\bibinfo {author} {\bibfnamefont {J.}~\bibnamefont
  {Li}}, \bibinfo {author} {\bibfnamefont {M.}~\bibnamefont {Nilsing}},
  \bibinfo {author} {\bibfnamefont {I.}~\bibnamefont {Kondov}}, \bibinfo
  {author} {\bibfnamefont {H.}~\bibnamefont {Wang}}, \bibinfo {author}
  {\bibfnamefont {P.}~\bibnamefont {Persson}}, \bibinfo {author} {\bibfnamefont
  {S.}~\bibnamefont {Lunell}},\ and\ \bibinfo {author} {\bibfnamefont
  {M.}~\bibnamefont {Thoss}},\ }\href {https://doi.org/10.1021/jp7118263}
  {\bibfield  {journal} {\bibinfo  {journal} {J. Phys. Chem. C}\ }\textbf
  {\bibinfo {volume} {112}},\ \bibinfo {pages} {12326} (\bibinfo {year}
  {2008})}\BibitemShut {NoStop}%
\bibitem [{\citenamefont {Li}\ \emph {et~al.}(2010)\citenamefont {Li},
  \citenamefont {Kondov}, \citenamefont {Wang},\ and\ \citenamefont
  {Thoss}}]{Jingrui2010}%
  \BibitemOpen
  \bibfield  {author} {\bibinfo {author} {\bibfnamefont {J.}~\bibnamefont
  {Li}}, \bibinfo {author} {\bibfnamefont {I.}~\bibnamefont {Kondov}}, \bibinfo
  {author} {\bibfnamefont {H.}~\bibnamefont {Wang}},\ and\ \bibinfo {author}
  {\bibfnamefont {M.}~\bibnamefont {Thoss}},\ }\href
  {https://doi.org/10.1021/jp104335k} {\bibfield  {journal} {\bibinfo
  {journal} {J. Phys. Chem. C}\ }\textbf {\bibinfo {volume} {114}},\ \bibinfo
  {pages} {18481} (\bibinfo {year} {2010})}\BibitemShut {NoStop}%
\bibitem [{\citenamefont {Li}\ \emph {et~al.}(2012)\citenamefont {Li},
  \citenamefont {Wang}, \citenamefont {Persson},\ and\ \citenamefont
  {Thoss}}]{Jingrui2012}%
  \BibitemOpen
  \bibfield  {author} {\bibinfo {author} {\bibfnamefont {J.}~\bibnamefont
  {Li}}, \bibinfo {author} {\bibfnamefont {H.}~\bibnamefont {Wang}}, \bibinfo
  {author} {\bibfnamefont {P.}~\bibnamefont {Persson}},\ and\ \bibinfo {author}
  {\bibfnamefont {M.}~\bibnamefont {Thoss}},\ }\href
  {https://doi.org/10.1063/1.4746768} {\bibfield  {journal} {\bibinfo
  {journal} {J. Chem. Phys.}\ }\textbf {\bibinfo {volume} {137}},\ \bibinfo
  {pages} {22A529} (\bibinfo {year} {2012})}\BibitemShut {NoStop}%
\bibitem [{\citenamefont {Li}\ \emph {et~al.}(2015)\citenamefont {Li},
  \citenamefont {Li}, \citenamefont {Winget},\ and\ \citenamefont
  {Bredas}}]{Jingrui2015}%
  \BibitemOpen
  \bibfield  {author} {\bibinfo {author} {\bibfnamefont {J.}~\bibnamefont
  {Li}}, \bibinfo {author} {\bibfnamefont {H.}~\bibnamefont {Li}}, \bibinfo
  {author} {\bibfnamefont {P.}~\bibnamefont {Winget}},\ and\ \bibinfo {author}
  {\bibfnamefont {J.-L.}\ \bibnamefont {Bredas}},\ }\href
  {https://doi.org/10.1021/acs.jpcc.5b03596} {\bibfield  {journal} {\bibinfo
  {journal} {J. Phys. Chem. C}\ }\textbf {\bibinfo {volume} {119}},\ \bibinfo
  {pages} {18843} (\bibinfo {year} {2015})}\BibitemShut {NoStop}%
\bibitem [{\citenamefont {Futera}\ and\ \citenamefont
  {Blumberger}(2017)}]{Futera2017}%
  \BibitemOpen
  \bibfield  {author} {\bibinfo {author} {\bibfnamefont {Z.}~\bibnamefont
  {Futera}}\ and\ \bibinfo {author} {\bibfnamefont {J.}~\bibnamefont
  {Blumberger}},\ }\href {https://doi.org/10.1021/acs.jpcc.7b06566} {\bibfield
  {journal} {\bibinfo  {journal} {J. Phys. Chem. C}\ }\textbf {\bibinfo
  {volume} {121}},\ \bibinfo {pages} {19677} (\bibinfo {year}
  {2017})}\BibitemShut {NoStop}%
\bibitem [{\citenamefont {Prucker}\ \emph {et~al.}(2013)\citenamefont
  {Prucker}, \citenamefont {Rubio-Pons}, \citenamefont {Bockstedte},
  \citenamefont {Wang}, \citenamefont {Coto},\ and\ \citenamefont
  {Thoss}}]{Prucker2013}%
  \BibitemOpen
  \bibfield  {author} {\bibinfo {author} {\bibfnamefont {V.}~\bibnamefont
  {Prucker}}, \bibinfo {author} {\bibfnamefont {O.}~\bibnamefont {Rubio-Pons}},
  \bibinfo {author} {\bibfnamefont {M.}~\bibnamefont {Bockstedte}}, \bibinfo
  {author} {\bibfnamefont {H.}~\bibnamefont {Wang}}, \bibinfo {author}
  {\bibfnamefont {P.~B.}\ \bibnamefont {Coto}},\ and\ \bibinfo {author}
  {\bibfnamefont {M.}~\bibnamefont {Thoss}},\ }\href
  {https://doi.org/10.1021/jp4091848} {\bibfield  {journal} {\bibinfo
  {journal} {J. Phys. Chem. C}\ }\textbf {\bibinfo {volume} {117}},\ \bibinfo
  {pages} {25334} (\bibinfo {year} {2013})}\BibitemShut {NoStop}%
\bibitem [{\citenamefont {Prucker}\ \emph {et~al.}(2018)\citenamefont
  {Prucker}, \citenamefont {Bockstedte}, \citenamefont {Thoss},\ and\
  \citenamefont {Coto}}]{Prucker2018}%
  \BibitemOpen
  \bibfield  {author} {\bibinfo {author} {\bibfnamefont {V.}~\bibnamefont
  {Prucker}}, \bibinfo {author} {\bibfnamefont {M.}~\bibnamefont {Bockstedte}},
  \bibinfo {author} {\bibfnamefont {M.}~\bibnamefont {Thoss}},\ and\ \bibinfo
  {author} {\bibfnamefont {P.~B.}\ \bibnamefont {Coto}},\ }\href
  {https://doi.org/10.1063/1.5020238} {\bibfield  {journal} {\bibinfo
  {journal} {J. Chem. Phys.}\ }\textbf {\bibinfo {volume} {148}},\ \bibinfo
  {pages} {124705} (\bibinfo {year} {2018})}\BibitemShut {NoStop}%
\bibitem [{\citenamefont {Pastore}\ and\ \citenamefont
  {De~Angelis}(2015)}]{Pastore2015}%
  \BibitemOpen
  \bibfield  {author} {\bibinfo {author} {\bibfnamefont {M.}~\bibnamefont
  {Pastore}}\ and\ \bibinfo {author} {\bibfnamefont {F.}~\bibnamefont
  {De~Angelis}},\ }\href {https://doi.org/10.1021/jacs.5b02128} {\bibfield
  {journal} {\bibinfo  {journal} {J. Am. Chem. Soc.}\ }\textbf {\bibinfo
  {volume} {137}},\ \bibinfo {pages} {5798} (\bibinfo {year}
  {2015})}\BibitemShut {NoStop}%
\bibitem [{\citenamefont {Warburton}\ \emph {et~al.}(2022)\citenamefont
  {Warburton}, \citenamefont {Soudackov},\ and\ \citenamefont
  {Hammes-Schiffer}}]{Schiffer2022}%
  \BibitemOpen
  \bibfield  {author} {\bibinfo {author} {\bibfnamefont {R.~E.}\ \bibnamefont
  {Warburton}}, \bibinfo {author} {\bibfnamefont {A.~V.}\ \bibnamefont
  {Soudackov}},\ and\ \bibinfo {author} {\bibfnamefont {S.}~\bibnamefont
  {Hammes-Schiffer}},\ }\href {https://doi.org/10.1021/acs.chemrev.1c00929}
  {\bibfield  {journal} {\bibinfo  {journal} {Chemical Reviews}\ }\textbf
  {\bibinfo {volume} {122}},\ \bibinfo {pages} {10599} (\bibinfo {year}
  {2022})}\BibitemShut {NoStop}%
\bibitem [{\citenamefont {Bahlke}\ \emph {et~al.}(2021)\citenamefont {Bahlke},
  \citenamefont {Schneeberger},\ and\ \citenamefont {Herrmann}}]{Bahlke2021}%
  \BibitemOpen
  \bibfield  {author} {\bibinfo {author} {\bibfnamefont {M.~P.}\ \bibnamefont
  {Bahlke}}, \bibinfo {author} {\bibfnamefont {M.}~\bibnamefont
  {Schneeberger}},\ and\ \bibinfo {author} {\bibfnamefont {C.}~\bibnamefont
  {Herrmann}},\ }\href {https://doi.org/10.1063/5.0045640} {\bibfield
  {journal} {\bibinfo  {journal} {J. Chem. Phys.}\ }\textbf {\bibinfo {volume}
  {154}},\ \bibinfo {pages} {144108} (\bibinfo {year} {2021})}\BibitemShut
  {NoStop}%
\bibitem [{\citenamefont {Kittel}\ and\ \citenamefont
  {Fong}(1987)}]{Kittel1987quantum}%
  \BibitemOpen
  \bibfield  {author} {\bibinfo {author} {\bibfnamefont {C.}~\bibnamefont
  {Kittel}}\ and\ \bibinfo {author} {\bibfnamefont {C.-y.}\ \bibnamefont
  {Fong}},\ }\href
  {https://www.wiley.com/en-us/Quantum+Theory+of+Solids\%2C+2nd+Revised+Edition-p-9780471624127}
  {\emph {\bibinfo {title} {Quantum theory of solids}}}\ (\bibinfo  {publisher}
  {Wiley},\ \bibinfo {year} {1987})\BibitemShut {NoStop}%
\bibitem [{\citenamefont {Gosavi}\ and\ \citenamefont
  {Marcus}(2000)}]{Marcus2000}%
  \BibitemOpen
  \bibfield  {author} {\bibinfo {author} {\bibfnamefont {S.}~\bibnamefont
  {Gosavi}}\ and\ \bibinfo {author} {\bibfnamefont {R.~A.}\ \bibnamefont
  {Marcus}},\ }\href {https://doi.org/10.1021/jp9933673} {\bibfield  {journal}
  {\bibinfo  {journal} {J. Phys. Chem. B}\ }\textbf {\bibinfo {volume} {104}},\
  \bibinfo {pages} {2067} (\bibinfo {year} {2000})}\BibitemShut {NoStop}%
\bibitem [{\citenamefont {Monkhorst}\ and\ \citenamefont
  {Pack}(1976)}]{Monkhorst1976}%
  \BibitemOpen
  \bibfield  {author} {\bibinfo {author} {\bibfnamefont {H.~J.}\ \bibnamefont
  {Monkhorst}}\ and\ \bibinfo {author} {\bibfnamefont {J.~D.}\ \bibnamefont
  {Pack}},\ }\href {https://doi.org/10.1103/PhysRevB.13.5188} {\bibfield
  {journal} {\bibinfo  {journal} {Phys. Rev. B}\ }\textbf {\bibinfo {volume}
  {13}},\ \bibinfo {pages} {5188} (\bibinfo {year} {1976})}\BibitemShut
  {NoStop}%
\bibitem [{\citenamefont {Valeev}\ \emph {et~al.}(2006)\citenamefont {Valeev},
  \citenamefont {Coropceanu}, \citenamefont {da~Silva~Filho}, \citenamefont
  {Salman},\ and\ \citenamefont {Brédas}}]{Valeev2006}%
  \BibitemOpen
  \bibfield  {author} {\bibinfo {author} {\bibfnamefont {E.~F.}\ \bibnamefont
  {Valeev}}, \bibinfo {author} {\bibfnamefont {V.}~\bibnamefont {Coropceanu}},
  \bibinfo {author} {\bibfnamefont {D.~A.}\ \bibnamefont {da~Silva~Filho}},
  \bibinfo {author} {\bibfnamefont {S.}~\bibnamefont {Salman}},\ and\ \bibinfo
  {author} {\bibfnamefont {J.-L.}\ \bibnamefont {Brédas}},\ }\href
  {https://doi.org/10.1021/ja061827h} {\bibfield  {journal} {\bibinfo
  {journal} {J. Am. Chem. Soc.}\ }\textbf {\bibinfo {volume} {128}},\ \bibinfo
  {pages} {9882} (\bibinfo {year} {2006})}\BibitemShut {NoStop}%
\bibitem [{\citenamefont {Baumeier}\ \emph {et~al.}(2010)\citenamefont
  {Baumeier}, \citenamefont {Kirkpatrick},\ and\ \citenamefont
  {Andrienko}}]{Baumeier2010}%
  \BibitemOpen
  \bibfield  {author} {\bibinfo {author} {\bibfnamefont {B.}~\bibnamefont
  {Baumeier}}, \bibinfo {author} {\bibfnamefont {J.}~\bibnamefont
  {Kirkpatrick}},\ and\ \bibinfo {author} {\bibfnamefont {D.}~\bibnamefont
  {Andrienko}},\ }\href {https://doi.org/10.1039/C002337J} {\bibfield
  {journal} {\bibinfo  {journal} {Phys. Chem. Chem. Phys.}\ }\textbf {\bibinfo
  {volume} {12}},\ \bibinfo {pages} {11103} (\bibinfo {year}
  {2010})}\BibitemShut {NoStop}%
\bibitem [{\citenamefont {Senthilkumar}\ \emph {et~al.}(2003)\citenamefont
  {Senthilkumar}, \citenamefont {Grozema}, \citenamefont {Bickelhaupt},\ and\
  \citenamefont {Siebbeles}}]{Siebbeles2003}%
  \BibitemOpen
  \bibfield  {author} {\bibinfo {author} {\bibfnamefont {K.}~\bibnamefont
  {Senthilkumar}}, \bibinfo {author} {\bibfnamefont {F.~C.}\ \bibnamefont
  {Grozema}}, \bibinfo {author} {\bibfnamefont {F.~M.}\ \bibnamefont
  {Bickelhaupt}},\ and\ \bibinfo {author} {\bibfnamefont {L.~D.~A.}\
  \bibnamefont {Siebbeles}},\ }\href {https://doi.org/10.1063/1.1615476}
  {\bibfield  {journal} {\bibinfo  {journal} {J. Chem. Phys.}\ }\textbf
  {\bibinfo {volume} {119}},\ \bibinfo {pages} {9809} (\bibinfo {year}
  {2003})}\BibitemShut {NoStop}%
\bibitem [{\citenamefont {Li}\ \emph {et~al.}(2007)\citenamefont {Li},
  \citenamefont {Brédas},\ and\ \citenamefont {Lennartz}}]{Li2007JCP}%
  \BibitemOpen
  \bibfield  {author} {\bibinfo {author} {\bibfnamefont {H.}~\bibnamefont
  {Li}}, \bibinfo {author} {\bibfnamefont {J.-L.}\ \bibnamefont {Brédas}},\
  and\ \bibinfo {author} {\bibfnamefont {C.}~\bibnamefont {Lennartz}},\ }\href
  {https://doi.org/10.1063/1.2727480} {\bibfield  {journal} {\bibinfo
  {journal} {J. Chem. Phys.}\ }\textbf {\bibinfo {volume} {126}},\ \bibinfo
  {pages} {164704} (\bibinfo {year} {2007})}\BibitemShut {NoStop}%
\bibitem [{\citenamefont {Wesolowski}\ and\ \citenamefont
  {Warshel}(1993)}]{Wesolowski1993JPC}%
  \BibitemOpen
  \bibfield  {author} {\bibinfo {author} {\bibfnamefont {T.~A.}\ \bibnamefont
  {Wesolowski}}\ and\ \bibinfo {author} {\bibfnamefont {A.}~\bibnamefont
  {Warshel}},\ }\href {https://doi.org/10.1021/j100132a040} {\bibfield
  {journal} {\bibinfo  {journal} {J. Phys. Chem.}\ }\textbf {\bibinfo {volume}
  {97}},\ \bibinfo {pages} {8050} (\bibinfo {year} {1993})}\BibitemShut
  {NoStop}%
\bibitem [{\citenamefont {Pavanello}\ and\ \citenamefont
  {Neugebauer}(2011)}]{Pavanello2011JCP}%
  \BibitemOpen
  \bibfield  {author} {\bibinfo {author} {\bibfnamefont {M.}~\bibnamefont
  {Pavanello}}\ and\ \bibinfo {author} {\bibfnamefont {J.}~\bibnamefont
  {Neugebauer}},\ }\href {https://doi.org/10.1063/1.3666005} {\bibfield
  {journal} {\bibinfo  {journal} {J. Chem. Phys.}\ }\textbf {\bibinfo {volume}
  {135}},\ \bibinfo {pages} {234103} (\bibinfo {year} {2011})}\BibitemShut
  {NoStop}%
\bibitem [{\citenamefont {Schober}\ \emph {et~al.}(2016)\citenamefont
  {Schober}, \citenamefont {Reuter},\ and\ \citenamefont
  {Oberhofer}}]{Schober2016}%
  \BibitemOpen
  \bibfield  {author} {\bibinfo {author} {\bibfnamefont {C.}~\bibnamefont
  {Schober}}, \bibinfo {author} {\bibfnamefont {K.}~\bibnamefont {Reuter}},\
  and\ \bibinfo {author} {\bibfnamefont {H.}~\bibnamefont {Oberhofer}},\ }\href
  {https://doi.org/10.1063/1.4940920} {\bibfield  {journal} {\bibinfo
  {journal} {J. Chem. Phys.}\ }\textbf {\bibinfo {volume} {144}},\ \bibinfo
  {pages} {054103} (\bibinfo {year} {2016})}\BibitemShut {NoStop}%
\bibitem [{\citenamefont {Gajdos}\ \emph {et~al.}(2014)\citenamefont {Gajdos},
  \citenamefont {Valner}, \citenamefont {Hoffmann}, \citenamefont {Spencer},
  \citenamefont {Breuer}, \citenamefont {Kubas}, \citenamefont {Dupuis},\ and\
  \citenamefont {Blumberger}}]{Gajdos2014JCTC}%
  \BibitemOpen
  \bibfield  {author} {\bibinfo {author} {\bibfnamefont {F.}~\bibnamefont
  {Gajdos}}, \bibinfo {author} {\bibfnamefont {S.}~\bibnamefont {Valner}},
  \bibinfo {author} {\bibfnamefont {F.}~\bibnamefont {Hoffmann}}, \bibinfo
  {author} {\bibfnamefont {J.}~\bibnamefont {Spencer}}, \bibinfo {author}
  {\bibfnamefont {M.}~\bibnamefont {Breuer}}, \bibinfo {author} {\bibfnamefont
  {A.}~\bibnamefont {Kubas}}, \bibinfo {author} {\bibfnamefont
  {M.}~\bibnamefont {Dupuis}},\ and\ \bibinfo {author} {\bibfnamefont
  {J.}~\bibnamefont {Blumberger}},\ }\href {https://doi.org/10.1021/ct500527v}
  {\bibfield  {journal} {\bibinfo  {journal} {J. Chem. Theory Comput.}\
  }\textbf {\bibinfo {volume} {10}},\ \bibinfo {pages} {4653} (\bibinfo {year}
  {2014})}\BibitemShut {NoStop}%
\bibitem [{\citenamefont {Fermi}(1950)}]{Fermi1950nuclear}%
  \BibitemOpen
  \bibfield  {author} {\bibinfo {author} {\bibfnamefont {E.}~\bibnamefont
  {Fermi}},\ }\href@noop {} {\emph {\bibinfo {title} {Nuclear physics: a course
  given by Enrico Fermi at the University of Chicago}}}\ (\bibinfo  {publisher}
  {University of Chicago Press},\ \bibinfo {year} {1950})\BibitemShut {NoStop}%
\bibitem [{\citenamefont {Sánchez-Portal}(2007)}]{Sanchez2007}%
  \BibitemOpen
  \bibfield  {author} {\bibinfo {author} {\bibfnamefont {D.}~\bibnamefont
  {Sánchez-Portal}},\ }\href {https://doi.org/10.1016/j.progsurf.2007.03.008}
  {\bibfield  {journal} {\bibinfo  {journal} {Prog. Surf. Sci.}\ }\textbf
  {\bibinfo {volume} {82}},\ \bibinfo {pages} {313} (\bibinfo {year}
  {2007})}\BibitemShut {NoStop}%
\bibitem [{\citenamefont {Perdew}\ \emph {et~al.}(1996)\citenamefont {Perdew},
  \citenamefont {Burke},\ and\ \citenamefont {Ernzerhof}}]{PBE}%
  \BibitemOpen
  \bibfield  {author} {\bibinfo {author} {\bibfnamefont {J.~P.}\ \bibnamefont
  {Perdew}}, \bibinfo {author} {\bibfnamefont {K.}~\bibnamefont {Burke}},\ and\
  \bibinfo {author} {\bibfnamefont {M.}~\bibnamefont {Ernzerhof}},\ }\href
  {https://doi.org/10.1103/PhysRevLett.77.3865} {\bibfield  {journal} {\bibinfo
   {journal} {Phys. Rev. Lett.}\ }\textbf {\bibinfo {volume} {77}},\ \bibinfo
  {pages} {3865} (\bibinfo {year} {1996})}\BibitemShut {NoStop}%
\bibitem [{\citenamefont {Tkatchenko}\ and\ \citenamefont
  {Scheffler}(2009)}]{vdW}%
  \BibitemOpen
  \bibfield  {author} {\bibinfo {author} {\bibfnamefont {A.}~\bibnamefont
  {Tkatchenko}}\ and\ \bibinfo {author} {\bibfnamefont {M.}~\bibnamefont
  {Scheffler}},\ }\href {https://doi.org/10.1103/PhysRevLett.102.073005}
  {\bibfield  {journal} {\bibinfo  {journal} {Phys. Rev. Lett.}\ }\textbf
  {\bibinfo {volume} {102}},\ \bibinfo {pages} {073005} (\bibinfo {year}
  {2009})}\BibitemShut {NoStop}%
\bibitem [{\citenamefont {Blum}\ \emph {et~al.}(2009)\citenamefont {Blum},
  \citenamefont {Gehrke}, \citenamefont {Hanke}, \citenamefont {Havu},
  \citenamefont {Havu}, \citenamefont {Ren}, \citenamefont {Reuter},\ and\
  \citenamefont {Scheffler}}]{Aims2009}%
  \BibitemOpen
  \bibfield  {author} {\bibinfo {author} {\bibfnamefont {V.}~\bibnamefont
  {Blum}}, \bibinfo {author} {\bibfnamefont {R.}~\bibnamefont {Gehrke}},
  \bibinfo {author} {\bibfnamefont {F.}~\bibnamefont {Hanke}}, \bibinfo
  {author} {\bibfnamefont {P.}~\bibnamefont {Havu}}, \bibinfo {author}
  {\bibfnamefont {V.}~\bibnamefont {Havu}}, \bibinfo {author} {\bibfnamefont
  {X.}~\bibnamefont {Ren}}, \bibinfo {author} {\bibfnamefont {K.}~\bibnamefont
  {Reuter}},\ and\ \bibinfo {author} {\bibfnamefont {M.}~\bibnamefont
  {Scheffler}},\ }\href {https://doi.org/10.1016/j.cpc.2009.06.022} {\bibfield
  {journal} {\bibinfo  {journal} {Comput. Phys. Commun.}\ }\textbf {\bibinfo
  {volume} {180}},\ \bibinfo {pages} {2175 } (\bibinfo {year}
  {2009})}\BibitemShut {NoStop}%
\bibitem [{\citenamefont {Zhang}\ \emph {et~al.}(2013)\citenamefont {Zhang},
  \citenamefont {Ren}, \citenamefont {Rinke}, \citenamefont {Blum},\ and\
  \citenamefont {Scheffler}}]{Aims2013}%
  \BibitemOpen
  \bibfield  {author} {\bibinfo {author} {\bibfnamefont {I.~Y.}\ \bibnamefont
  {Zhang}}, \bibinfo {author} {\bibfnamefont {X.}~\bibnamefont {Ren}}, \bibinfo
  {author} {\bibfnamefont {P.}~\bibnamefont {Rinke}}, \bibinfo {author}
  {\bibfnamefont {V.}~\bibnamefont {Blum}},\ and\ \bibinfo {author}
  {\bibfnamefont {M.}~\bibnamefont {Scheffler}},\ }\href
  {https://doi.org/10.1088/1367-2630/15/12/123033} {\bibfield  {journal}
  {\bibinfo  {journal} {New Journal of Physics}\ }\textbf {\bibinfo {volume}
  {15}},\ \bibinfo {pages} {123033} (\bibinfo {year} {2013})}\BibitemShut
  {NoStop}%
\bibitem [{\citenamefont {Pritchard}\ \emph {et~al.}(2019)\citenamefont
  {Pritchard}, \citenamefont {Altarawy}, \citenamefont {Didier}, \citenamefont
  {Gibson},\ and\ \citenamefont {Windus}}]{BSE2019}%
  \BibitemOpen
  \bibfield  {author} {\bibinfo {author} {\bibfnamefont {B.~P.}\ \bibnamefont
  {Pritchard}}, \bibinfo {author} {\bibfnamefont {D.}~\bibnamefont {Altarawy}},
  \bibinfo {author} {\bibfnamefont {B.}~\bibnamefont {Didier}}, \bibinfo
  {author} {\bibfnamefont {T.~D.}\ \bibnamefont {Gibson}},\ and\ \bibinfo
  {author} {\bibfnamefont {T.~L.}\ \bibnamefont {Windus}},\ }\href
  {https://doi.org/10.1021/acs.jcim.9b00725} {\bibfield  {journal} {\bibinfo
  {journal} {J. Chem. Inf. Model.}\ }\textbf {\bibinfo {volume} {59}},\
  \bibinfo {pages} {4814} (\bibinfo {year} {2019})}\BibitemShut {NoStop}%
\bibitem [{\citenamefont {Michelitsch}\ and\ \citenamefont
  {Reuter}(2019)}]{Mitch_2019}%
  \BibitemOpen
  \bibfield  {author} {\bibinfo {author} {\bibfnamefont {G.~S.}\ \bibnamefont
  {Michelitsch}}\ and\ \bibinfo {author} {\bibfnamefont {K.}~\bibnamefont
  {Reuter}},\ }\href {https://doi.org/10.1063/1.5083618} {\bibfield  {journal}
  {\bibinfo  {journal} {J. Chem. Phys.}\ }\textbf {\bibinfo {volume} {150}},\
  \bibinfo {pages} {074104} (\bibinfo {year} {2019})}\BibitemShut {NoStop}%
\bibitem [{\citenamefont {Leetmaa}\ \emph {et~al.}(2010)\citenamefont
  {Leetmaa}, \citenamefont {Ljungberg}, \citenamefont {Lyubartsev},
  \citenamefont {Nilsson},\ and\ \citenamefont {Pettersson}}]{Leetmaa_2010}%
  \BibitemOpen
  \bibfield  {author} {\bibinfo {author} {\bibfnamefont {M.}~\bibnamefont
  {Leetmaa}}, \bibinfo {author} {\bibfnamefont {M.}~\bibnamefont {Ljungberg}},
  \bibinfo {author} {\bibfnamefont {A.}~\bibnamefont {Lyubartsev}}, \bibinfo
  {author} {\bibfnamefont {A.}~\bibnamefont {Nilsson}},\ and\ \bibinfo {author}
  {\bibfnamefont {L.}~\bibnamefont {Pettersson}},\ }\href
  {https://doi.org/10.1016/j.elspec.2010.02.004} {\bibfield  {journal}
  {\bibinfo  {journal} {J. Electron Spectrosc. Relat. Phenom.}\ }\textbf
  {\bibinfo {volume} {177}},\ \bibinfo {pages} {135} (\bibinfo {year}
  {2010})}\BibitemShut {NoStop}%
\bibitem [{\citenamefont {Virtanen}\ \emph {et~al.}(2020)\citenamefont
  {Virtanen}, \citenamefont {Gommers}, \citenamefont {Oliphant}, \citenamefont
  {Haberland}, \citenamefont {Reddy}, \citenamefont {Cournapeau}, \citenamefont
  {Burovski}, \citenamefont {Peterson}, \citenamefont {Weckesser},
  \citenamefont {Bright}, \citenamefont {van~der Walt}, \citenamefont {Brett},
  \citenamefont {Wilson}, \citenamefont {Millman}, \citenamefont {Mayorov},
  \citenamefont {Nelson}, \citenamefont {Jones}, \citenamefont {Kern},
  \citenamefont {Larson}, \citenamefont {Carey}, \citenamefont {Polat},
  \citenamefont {Feng}, \citenamefont {Moore}, \citenamefont {VanderPlas},
  \citenamefont {Laxalde}, \citenamefont {Perktold}, \citenamefont {Cimrman},
  \citenamefont {Henriksen}, \citenamefont {Quintero}, \citenamefont {Harris},
  \citenamefont {Archibald}, \citenamefont {Ribeiro}, \citenamefont
  {Pedregosa}, \citenamefont {van Mulbregt}, \citenamefont {Vijaykumar},
  \citenamefont {Bardelli}, \citenamefont {Rothberg}, \citenamefont {Hilboll},
  \citenamefont {Kloeckner}, \citenamefont {Scopatz}, \citenamefont {Lee},
  \citenamefont {Rokem}, \citenamefont {Woods}, \citenamefont {Fulton},
  \citenamefont {Masson}, \citenamefont {H{\"a}ggstr{\"o}m}, \citenamefont
  {Fitzgerald}, \citenamefont {Nicholson}, \citenamefont {Hagen}, \citenamefont
  {Pasechnik}, \citenamefont {Olivetti}, \citenamefont {Martin}, \citenamefont
  {Wieser}, \citenamefont {Silva}, \citenamefont {Lenders}, \citenamefont
  {Wilhelm}, \citenamefont {Young}, \citenamefont {Price}, \citenamefont
  {Ingold}, \citenamefont {Allen}, \citenamefont {Lee}, \citenamefont {Audren},
  \citenamefont {Probst}, \citenamefont {Dietrich}, \citenamefont {Silterra},
  \citenamefont {Webber}, \citenamefont {Slavi{\v{c}}}, \citenamefont
  {Nothman}, \citenamefont {Buchner}, \citenamefont {Kulick}, \citenamefont
  {Sch{\"o}nberger}, \citenamefont {de~Miranda~Cardoso}, \citenamefont
  {Reimer}, \citenamefont {Harrington}, \citenamefont {Rodr{\'i}guez},
  \citenamefont {Nunez-Iglesias}, \citenamefont {Kuczynski}, \citenamefont
  {Tritz}, \citenamefont {Thoma}, \citenamefont {Newville}, \citenamefont
  {K{\"u}mmerer}, \citenamefont {Bolingbroke}, \citenamefont {Tartre},
  \citenamefont {Pak}, \citenamefont {Smith}, \citenamefont {Nowaczyk},
  \citenamefont {Shebanov}, \citenamefont {Pavlyk}, \citenamefont {Brodtkorb},
  \citenamefont {Lee}, \citenamefont {McGibbon}, \citenamefont {Feldbauer},
  \citenamefont {Lewis}, \citenamefont {Tygier}, \citenamefont {Sievert},
  \citenamefont {Vigna}, \citenamefont {Peterson}, \citenamefont {More},
  \citenamefont {Pudlik}, \citenamefont {Oshima}, \citenamefont {Pingel},
  \citenamefont {Robitaille}, \citenamefont {Spura}, \citenamefont {Jones},
  \citenamefont {Cera}, \citenamefont {Leslie}, \citenamefont {Zito},
  \citenamefont {Krauss}, \citenamefont {Upadhyay}, \citenamefont {Halchenko},\
  and\ \citenamefont {V{\'a}zquez-Baeza}}]{SciPy}%
  \BibitemOpen
  \bibfield  {author} {\bibinfo {author} {\bibfnamefont {P.}~\bibnamefont
  {Virtanen}}, \bibinfo {author} {\bibfnamefont {R.}~\bibnamefont {Gommers}},
  \bibinfo {author} {\bibfnamefont {T.~E.}\ \bibnamefont {Oliphant}}, \bibinfo
  {author} {\bibfnamefont {M.}~\bibnamefont {Haberland}}, \bibinfo {author}
  {\bibfnamefont {T.}~\bibnamefont {Reddy}}, \bibinfo {author} {\bibfnamefont
  {D.}~\bibnamefont {Cournapeau}}, \bibinfo {author} {\bibfnamefont
  {E.}~\bibnamefont {Burovski}}, \bibinfo {author} {\bibfnamefont
  {P.}~\bibnamefont {Peterson}}, \bibinfo {author} {\bibfnamefont
  {W.}~\bibnamefont {Weckesser}}, \bibinfo {author} {\bibfnamefont
  {J.}~\bibnamefont {Bright}}, \bibinfo {author} {\bibfnamefont {S.~J.}\
  \bibnamefont {van~der Walt}}, \bibinfo {author} {\bibfnamefont
  {M.}~\bibnamefont {Brett}}, \bibinfo {author} {\bibfnamefont
  {J.}~\bibnamefont {Wilson}}, \bibinfo {author} {\bibfnamefont {K.~J.}\
  \bibnamefont {Millman}}, \bibinfo {author} {\bibfnamefont {N.}~\bibnamefont
  {Mayorov}}, \bibinfo {author} {\bibfnamefont {A.~R.~J.}\ \bibnamefont
  {Nelson}}, \bibinfo {author} {\bibfnamefont {E.}~\bibnamefont {Jones}},
  \bibinfo {author} {\bibfnamefont {R.}~\bibnamefont {Kern}}, \bibinfo {author}
  {\bibfnamefont {E.}~\bibnamefont {Larson}}, \bibinfo {author} {\bibfnamefont
  {C.~J.}\ \bibnamefont {Carey}}, \bibinfo {author} {\bibfnamefont
  {{\.{I}}.}~\bibnamefont {Polat}}, \bibinfo {author} {\bibfnamefont
  {Y.}~\bibnamefont {Feng}}, \bibinfo {author} {\bibfnamefont {E.~W.}\
  \bibnamefont {Moore}}, \bibinfo {author} {\bibfnamefont {J.}~\bibnamefont
  {VanderPlas}}, \bibinfo {author} {\bibfnamefont {D.}~\bibnamefont {Laxalde}},
  \bibinfo {author} {\bibfnamefont {J.}~\bibnamefont {Perktold}}, \bibinfo
  {author} {\bibfnamefont {R.}~\bibnamefont {Cimrman}}, \bibinfo {author}
  {\bibfnamefont {I.}~\bibnamefont {Henriksen}}, \bibinfo {author}
  {\bibfnamefont {E.~A.}\ \bibnamefont {Quintero}}, \bibinfo {author}
  {\bibfnamefont {C.~R.}\ \bibnamefont {Harris}}, \bibinfo {author}
  {\bibfnamefont {A.~M.}\ \bibnamefont {Archibald}}, \bibinfo {author}
  {\bibfnamefont {A.~H.}\ \bibnamefont {Ribeiro}}, \bibinfo {author}
  {\bibfnamefont {F.}~\bibnamefont {Pedregosa}}, \bibinfo {author}
  {\bibfnamefont {P.}~\bibnamefont {van Mulbregt}}, \bibinfo {author}
  {\bibfnamefont {A.}~\bibnamefont {Vijaykumar}}, \bibinfo {author}
  {\bibfnamefont {A.~P.}\ \bibnamefont {Bardelli}}, \bibinfo {author}
  {\bibfnamefont {A.}~\bibnamefont {Rothberg}}, \bibinfo {author}
  {\bibfnamefont {A.}~\bibnamefont {Hilboll}}, \bibinfo {author} {\bibfnamefont
  {A.}~\bibnamefont {Kloeckner}}, \bibinfo {author} {\bibfnamefont
  {A.}~\bibnamefont {Scopatz}}, \bibinfo {author} {\bibfnamefont
  {A.}~\bibnamefont {Lee}}, \bibinfo {author} {\bibfnamefont {A.}~\bibnamefont
  {Rokem}}, \bibinfo {author} {\bibfnamefont {C.~N.}\ \bibnamefont {Woods}},
  \bibinfo {author} {\bibfnamefont {C.}~\bibnamefont {Fulton}}, \bibinfo
  {author} {\bibfnamefont {C.}~\bibnamefont {Masson}}, \bibinfo {author}
  {\bibfnamefont {C.}~\bibnamefont {H{\"a}ggstr{\"o}m}}, \bibinfo {author}
  {\bibfnamefont {C.}~\bibnamefont {Fitzgerald}}, \bibinfo {author}
  {\bibfnamefont {D.~A.}\ \bibnamefont {Nicholson}}, \bibinfo {author}
  {\bibfnamefont {D.~R.}\ \bibnamefont {Hagen}}, \bibinfo {author}
  {\bibfnamefont {D.~V.}\ \bibnamefont {Pasechnik}}, \bibinfo {author}
  {\bibfnamefont {E.}~\bibnamefont {Olivetti}}, \bibinfo {author}
  {\bibfnamefont {E.}~\bibnamefont {Martin}}, \bibinfo {author} {\bibfnamefont
  {E.}~\bibnamefont {Wieser}}, \bibinfo {author} {\bibfnamefont
  {F.}~\bibnamefont {Silva}}, \bibinfo {author} {\bibfnamefont
  {F.}~\bibnamefont {Lenders}}, \bibinfo {author} {\bibfnamefont
  {F.}~\bibnamefont {Wilhelm}}, \bibinfo {author} {\bibfnamefont
  {G.}~\bibnamefont {Young}}, \bibinfo {author} {\bibfnamefont {G.~A.}\
  \bibnamefont {Price}}, \bibinfo {author} {\bibfnamefont {G.-L.}\ \bibnamefont
  {Ingold}}, \bibinfo {author} {\bibfnamefont {G.~E.}\ \bibnamefont {Allen}},
  \bibinfo {author} {\bibfnamefont {G.~R.}\ \bibnamefont {Lee}}, \bibinfo
  {author} {\bibfnamefont {H.}~\bibnamefont {Audren}}, \bibinfo {author}
  {\bibfnamefont {I.}~\bibnamefont {Probst}}, \bibinfo {author} {\bibfnamefont
  {J.~P.}\ \bibnamefont {Dietrich}}, \bibinfo {author} {\bibfnamefont
  {J.}~\bibnamefont {Silterra}}, \bibinfo {author} {\bibfnamefont {J.~T.}\
  \bibnamefont {Webber}}, \bibinfo {author} {\bibfnamefont {J.}~\bibnamefont
  {Slavi{\v{c}}}}, \bibinfo {author} {\bibfnamefont {J.}~\bibnamefont
  {Nothman}}, \bibinfo {author} {\bibfnamefont {J.}~\bibnamefont {Buchner}},
  \bibinfo {author} {\bibfnamefont {J.}~\bibnamefont {Kulick}}, \bibinfo
  {author} {\bibfnamefont {J.~L.}\ \bibnamefont {Sch{\"o}nberger}}, \bibinfo
  {author} {\bibfnamefont {J.~V.}\ \bibnamefont {de~Miranda~Cardoso}}, \bibinfo
  {author} {\bibfnamefont {J.}~\bibnamefont {Reimer}}, \bibinfo {author}
  {\bibfnamefont {J.}~\bibnamefont {Harrington}}, \bibinfo {author}
  {\bibfnamefont {J.~L.~C.}\ \bibnamefont {Rodr{\'i}guez}}, \bibinfo {author}
  {\bibfnamefont {J.}~\bibnamefont {Nunez-Iglesias}}, \bibinfo {author}
  {\bibfnamefont {J.}~\bibnamefont {Kuczynski}}, \bibinfo {author}
  {\bibfnamefont {K.}~\bibnamefont {Tritz}}, \bibinfo {author} {\bibfnamefont
  {M.}~\bibnamefont {Thoma}}, \bibinfo {author} {\bibfnamefont
  {M.}~\bibnamefont {Newville}}, \bibinfo {author} {\bibfnamefont
  {M.}~\bibnamefont {K{\"u}mmerer}}, \bibinfo {author} {\bibfnamefont
  {M.}~\bibnamefont {Bolingbroke}}, \bibinfo {author} {\bibfnamefont
  {M.}~\bibnamefont {Tartre}}, \bibinfo {author} {\bibfnamefont
  {M.}~\bibnamefont {Pak}}, \bibinfo {author} {\bibfnamefont {N.~J.}\
  \bibnamefont {Smith}}, \bibinfo {author} {\bibfnamefont {N.}~\bibnamefont
  {Nowaczyk}}, \bibinfo {author} {\bibfnamefont {N.}~\bibnamefont {Shebanov}},
  \bibinfo {author} {\bibfnamefont {O.}~\bibnamefont {Pavlyk}}, \bibinfo
  {author} {\bibfnamefont {P.~A.}\ \bibnamefont {Brodtkorb}}, \bibinfo {author}
  {\bibfnamefont {P.}~\bibnamefont {Lee}}, \bibinfo {author} {\bibfnamefont
  {R.~T.}\ \bibnamefont {McGibbon}}, \bibinfo {author} {\bibfnamefont
  {R.}~\bibnamefont {Feldbauer}}, \bibinfo {author} {\bibfnamefont
  {S.}~\bibnamefont {Lewis}}, \bibinfo {author} {\bibfnamefont
  {S.}~\bibnamefont {Tygier}}, \bibinfo {author} {\bibfnamefont
  {S.}~\bibnamefont {Sievert}}, \bibinfo {author} {\bibfnamefont
  {S.}~\bibnamefont {Vigna}}, \bibinfo {author} {\bibfnamefont
  {S.}~\bibnamefont {Peterson}}, \bibinfo {author} {\bibfnamefont
  {S.}~\bibnamefont {More}}, \bibinfo {author} {\bibfnamefont {T.}~\bibnamefont
  {Pudlik}}, \bibinfo {author} {\bibfnamefont {T.}~\bibnamefont {Oshima}},
  \bibinfo {author} {\bibfnamefont {T.~J.}\ \bibnamefont {Pingel}}, \bibinfo
  {author} {\bibfnamefont {T.~P.}\ \bibnamefont {Robitaille}}, \bibinfo
  {author} {\bibfnamefont {T.}~\bibnamefont {Spura}}, \bibinfo {author}
  {\bibfnamefont {T.~R.}\ \bibnamefont {Jones}}, \bibinfo {author}
  {\bibfnamefont {T.}~\bibnamefont {Cera}}, \bibinfo {author} {\bibfnamefont
  {T.}~\bibnamefont {Leslie}}, \bibinfo {author} {\bibfnamefont
  {T.}~\bibnamefont {Zito}}, \bibinfo {author} {\bibfnamefont {T.}~\bibnamefont
  {Krauss}}, \bibinfo {author} {\bibfnamefont {U.}~\bibnamefont {Upadhyay}},
  \bibinfo {author} {\bibfnamefont {Y.~O.}\ \bibnamefont {Halchenko}},\ and\
  \bibinfo {author} {\bibfnamefont {Y.}~\bibnamefont {V{\'a}zquez-Baeza}},\
  }\href {https://doi.org/10.1038/s41592-019-0686-2} {\bibfield  {journal}
  {\bibinfo  {journal} {Nat. Methods.}\ }\textbf {\bibinfo {volume} {17}},\
  \bibinfo {pages} {261} (\bibinfo {year} {2020})}\BibitemShut {NoStop}%
\bibitem [{\citenamefont {Karis}\ \emph {et~al.}(1996)\citenamefont {Karis},
  \citenamefont {Nilsson}, \citenamefont {Weinelt}, \citenamefont {Wiell},
  \citenamefont {Puglia}, \citenamefont {Wassdahl}, \citenamefont
  {M\aa{}rtensson}, \citenamefont {Samant},\ and\ \citenamefont
  {St\"ohr}}]{Nilsson_1996}%
  \BibitemOpen
  \bibfield  {author} {\bibinfo {author} {\bibfnamefont {O.}~\bibnamefont
  {Karis}}, \bibinfo {author} {\bibfnamefont {A.}~\bibnamefont {Nilsson}},
  \bibinfo {author} {\bibfnamefont {M.}~\bibnamefont {Weinelt}}, \bibinfo
  {author} {\bibfnamefont {T.}~\bibnamefont {Wiell}}, \bibinfo {author}
  {\bibfnamefont {C.}~\bibnamefont {Puglia}}, \bibinfo {author} {\bibfnamefont
  {N.}~\bibnamefont {Wassdahl}}, \bibinfo {author} {\bibfnamefont
  {N.}~\bibnamefont {M\aa{}rtensson}}, \bibinfo {author} {\bibfnamefont
  {M.}~\bibnamefont {Samant}},\ and\ \bibinfo {author} {\bibfnamefont
  {J.}~\bibnamefont {St\"ohr}},\ }\href
  {https://doi.org/10.1103/PhysRevLett.76.1380} {\bibfield  {journal} {\bibinfo
   {journal} {Phys. Rev. Lett.}\ }\textbf {\bibinfo {volume} {76}},\ \bibinfo
  {pages} {1380} (\bibinfo {year} {1996})}\BibitemShut {NoStop}%
\bibitem [{\citenamefont {Keller}\ \emph
  {et~al.}(1998{\natexlab{b}})\citenamefont {Keller}, \citenamefont {Stichler},
  \citenamefont {Comelli}, \citenamefont {Esch}, \citenamefont {Lizzit},
  \citenamefont {Wurth},\ and\ \citenamefont {Menzel}}]{Keller1998}%
  \BibitemOpen
  \bibfield  {author} {\bibinfo {author} {\bibfnamefont {C.}~\bibnamefont
  {Keller}}, \bibinfo {author} {\bibfnamefont {M.}~\bibnamefont {Stichler}},
  \bibinfo {author} {\bibfnamefont {G.}~\bibnamefont {Comelli}}, \bibinfo
  {author} {\bibfnamefont {F.}~\bibnamefont {Esch}}, \bibinfo {author}
  {\bibfnamefont {S.}~\bibnamefont {Lizzit}}, \bibinfo {author} {\bibfnamefont
  {W.}~\bibnamefont {Wurth}},\ and\ \bibinfo {author} {\bibfnamefont
  {D.}~\bibnamefont {Menzel}},\ }\href
  {https://doi.org/10.1103/PhysRevLett.80.1774} {\bibfield  {journal} {\bibinfo
   {journal} {Phys. Rev. Lett.}\ }\textbf {\bibinfo {volume} {80}},\ \bibinfo
  {pages} {1774} (\bibinfo {year} {1998}{\natexlab{b}})}\BibitemShut {NoStop}%
\bibitem [{\citenamefont {Gross}\ \emph {et~al.}(2011)\citenamefont {Gross},
  \citenamefont {Moll}, \citenamefont {Mohn}, \citenamefont {Curioni},
  \citenamefont {Meyer}, \citenamefont {Hanke},\ and\ \citenamefont
  {Persson}}]{Gross2011}%
  \BibitemOpen
  \bibfield  {author} {\bibinfo {author} {\bibfnamefont {L.}~\bibnamefont
  {Gross}}, \bibinfo {author} {\bibfnamefont {N.}~\bibnamefont {Moll}},
  \bibinfo {author} {\bibfnamefont {F.}~\bibnamefont {Mohn}}, \bibinfo {author}
  {\bibfnamefont {A.}~\bibnamefont {Curioni}}, \bibinfo {author} {\bibfnamefont
  {G.}~\bibnamefont {Meyer}}, \bibinfo {author} {\bibfnamefont
  {F.}~\bibnamefont {Hanke}},\ and\ \bibinfo {author} {\bibfnamefont
  {M.}~\bibnamefont {Persson}},\ }\href
  {https://doi.org/10.1103/PhysRevLett.107.086101} {\bibfield  {journal}
  {\bibinfo  {journal} {Phys. Rev. Lett.}\ }\textbf {\bibinfo {volume} {107}},\
  \bibinfo {pages} {086101} (\bibinfo {year} {2011})}\BibitemShut {NoStop}%
\bibitem [{\citenamefont {Borisov}\ \emph {et~al.}(2002)\citenamefont
  {Borisov}, \citenamefont {Gauyacq}, \citenamefont {Chulkov}, \citenamefont
  {Silkin},\ and\ \citenamefont {Echenique}}]{Echenique2002}%
  \BibitemOpen
  \bibfield  {author} {\bibinfo {author} {\bibfnamefont {A.~G.}\ \bibnamefont
  {Borisov}}, \bibinfo {author} {\bibfnamefont {J.~P.}\ \bibnamefont
  {Gauyacq}}, \bibinfo {author} {\bibfnamefont {E.~V.}\ \bibnamefont
  {Chulkov}}, \bibinfo {author} {\bibfnamefont {V.~M.}\ \bibnamefont
  {Silkin}},\ and\ \bibinfo {author} {\bibfnamefont {P.~M.}\ \bibnamefont
  {Echenique}},\ }\href {https://doi.org/10.1103/PhysRevB.65.235434} {\bibfield
   {journal} {\bibinfo  {journal} {Phys. Rev. B}\ }\textbf {\bibinfo {volume}
  {65}},\ \bibinfo {pages} {235434} (\bibinfo {year} {2002})}\BibitemShut
  {NoStop}%
\bibitem [{\citenamefont {Repp}\ \emph {et~al.}(2005)\citenamefont {Repp},
  \citenamefont {Meyer}, \citenamefont {Stojkovi\ifmmode~\acute{c}\else
  \'{c}\fi{}}, \citenamefont {Gourdon},\ and\ \citenamefont
  {Joachim}}]{Repp2005}%
  \BibitemOpen
  \bibfield  {author} {\bibinfo {author} {\bibfnamefont {J.}~\bibnamefont
  {Repp}}, \bibinfo {author} {\bibfnamefont {G.}~\bibnamefont {Meyer}},
  \bibinfo {author} {\bibfnamefont {S.~M.}\ \bibnamefont
  {Stojkovi\ifmmode~\acute{c}\else \'{c}\fi{}}}, \bibinfo {author}
  {\bibfnamefont {A.}~\bibnamefont {Gourdon}},\ and\ \bibinfo {author}
  {\bibfnamefont {C.}~\bibnamefont {Joachim}},\ }\href
  {https://doi.org/10.1103/PhysRevLett.94.026803} {\bibfield  {journal}
  {\bibinfo  {journal} {Phys. Rev. Lett.}\ }\textbf {\bibinfo {volume} {94}},\
  \bibinfo {pages} {026803} (\bibinfo {year} {2005})}\BibitemShut {NoStop}%
\bibitem [{\citenamefont {Dou}\ and\ \citenamefont
  {Subotnik}(2016)}]{Subotnik2016}%
  \BibitemOpen
  \bibfield  {author} {\bibinfo {author} {\bibfnamefont {W.}~\bibnamefont
  {Dou}}\ and\ \bibinfo {author} {\bibfnamefont {J.~E.}\ \bibnamefont
  {Subotnik}},\ }\href {https://doi.org/10.1063/1.4959604} {\bibfield
  {journal} {\bibinfo  {journal} {J. Chem. Phys.}\ }\textbf {\bibinfo {volume}
  {145}},\ \bibinfo {pages} {054102} (\bibinfo {year} {2016})}\BibitemShut
  {NoStop}%
\bibitem [{\citenamefont {Dou}\ and\ \citenamefont
  {Subotnik}(2017)}]{Subotnik2017}%
  \BibitemOpen
  \bibfield  {author} {\bibinfo {author} {\bibfnamefont {W.}~\bibnamefont
  {Dou}}\ and\ \bibinfo {author} {\bibfnamefont {J.~E.}\ \bibnamefont
  {Subotnik}},\ }\href {https://doi.org/10.1063/1.4965823} {\bibfield
  {journal} {\bibinfo  {journal} {J. Chem. Phys.}\ }\textbf {\bibinfo {volume}
  {146}},\ \bibinfo {pages} {092304} (\bibinfo {year} {2017})}\BibitemShut
  {NoStop}%
\bibitem [{\citenamefont {Dou}\ and\ \citenamefont
  {Subotnik}(2018)}]{Subotnik2018}%
  \BibitemOpen
  \bibfield  {author} {\bibinfo {author} {\bibfnamefont {W.}~\bibnamefont
  {Dou}}\ and\ \bibinfo {author} {\bibfnamefont {J.~E.}\ \bibnamefont
  {Subotnik}},\ }\href {https://doi.org/10.1063/1.5035412} {\bibfield
  {journal} {\bibinfo  {journal} {J. Chem. Phys.}\ }\textbf {\bibinfo {volume}
  {148}},\ \bibinfo {pages} {230901} (\bibinfo {year} {2018})}\BibitemShut
  {NoStop}%
\bibitem [{\citenamefont {Cao}\ \emph {et~al.}(2022)\citenamefont {Cao},
  \citenamefont {Bukas}, \citenamefont {Shadravan}, \citenamefont {Wang},
  \citenamefont {Li}, \citenamefont {Kibsgaard}, \citenamefont {Chorkendorff},\
  and\ \citenamefont {N{\o}rskov}}]{Bukas2022}%
  \BibitemOpen
  \bibfield  {author} {\bibinfo {author} {\bibfnamefont {A.}~\bibnamefont
  {Cao}}, \bibinfo {author} {\bibfnamefont {V.~J.}\ \bibnamefont {Bukas}},
  \bibinfo {author} {\bibfnamefont {V.}~\bibnamefont {Shadravan}}, \bibinfo
  {author} {\bibfnamefont {Z.}~\bibnamefont {Wang}}, \bibinfo {author}
  {\bibfnamefont {H.}~\bibnamefont {Li}}, \bibinfo {author} {\bibfnamefont
  {J.}~\bibnamefont {Kibsgaard}}, \bibinfo {author} {\bibfnamefont
  {I.}~\bibnamefont {Chorkendorff}},\ and\ \bibinfo {author} {\bibfnamefont
  {J.~K.}\ \bibnamefont {N{\o}rskov}},\ }\href
  {https://doi.org/10.1038/s41467-022-30034-y} {\bibfield  {journal} {\bibinfo
  {journal} {Nat. Commun.}\ }\textbf {\bibinfo {volume} {13}},\ \bibinfo
  {pages} {2382} (\bibinfo {year} {2022})}\BibitemShut {NoStop}%
\bibitem [{\citenamefont {Vijay}\ \emph {et~al.}(2022)\citenamefont {Vijay},
  \citenamefont {Kastlunger}, \citenamefont {Chan},\ and\ \citenamefont
  {Nørskov}}]{Kastlunger2022}%
  \BibitemOpen
  \bibfield  {author} {\bibinfo {author} {\bibfnamefont {S.}~\bibnamefont
  {Vijay}}, \bibinfo {author} {\bibfnamefont {G.}~\bibnamefont {Kastlunger}},
  \bibinfo {author} {\bibfnamefont {K.}~\bibnamefont {Chan}},\ and\ \bibinfo
  {author} {\bibfnamefont {J.~K.}\ \bibnamefont {Nørskov}},\ }\href
  {https://doi.org/10.1063/5.0096625} {\bibfield  {journal} {\bibinfo
  {journal} {J. Chem. Phys.}\ }\textbf {\bibinfo {volume} {156}},\ \bibinfo
  {pages} {231102} (\bibinfo {year} {2022})}\BibitemShut {NoStop}%
\bibitem [{\citenamefont {Bhattacharjee}\ \emph {et~al.}(2016)\citenamefont
  {Bhattacharjee}, \citenamefont {Waghmare},\ and\ \citenamefont
  {Lee}}]{Bhattacharjee2016}%
  \BibitemOpen
  \bibfield  {author} {\bibinfo {author} {\bibfnamefont {S.}~\bibnamefont
  {Bhattacharjee}}, \bibinfo {author} {\bibfnamefont {U.~V.}\ \bibnamefont
  {Waghmare}},\ and\ \bibinfo {author} {\bibfnamefont {S.-C.}\ \bibnamefont
  {Lee}},\ }\href {https://doi.org/10.1038/srep35916} {\bibfield  {journal}
  {\bibinfo  {journal} {Sci. Rep.}\ }\textbf {\bibinfo {volume} {6}},\ \bibinfo
  {pages} {35916} (\bibinfo {year} {2016})}\BibitemShut {NoStop}%
\end{thebibliography}%

\end{document}